\pdfoutput=1
\documentclass[usegraphicx,useAMS,usenatbib]{mn2e}
\usepackage{graphicx,rotating,amsmath,hyperref}
\usepackage{tikz,amssymb}
\usepackage{MnSymbol}
\usepackage{fixltx2e}
\usepackage{gensymb}
\usepackage{subcaption}
\usepackage{booktabs}

\newcommand{\kms}{\mbox{km\,s$^{-1}$}}
\def\kmsn{km\,${\rm s}^{-1}$}
\def\mss{m\,${\rm s}^{-1}$}
\newcommand{\subs}[1]{$_{\rm #1}$}

\def\vsini{$\!${\em v\,}sin{\em i}}
\def\vsinis{$\!${\em v\,}sin{\em i} }
\def\ga{\mathrel{\hbox{\rlap{\hbox{\lower4pt\hbox{$\sim$}}}\hbox{$>$}}}}
\def\la{\mathrel{\hbox{\rlap{\hbox{\lower4pt\hbox{$\sim$}}}\hbox{$<$}}}}
\def\rdd{$\rm{rad\,{d}^{-1}}$}

\title[Magnetic Fields on Moderate Rotators: EK~Draconis]{Magnetic fields on young, moderately rotating Sun-like stars II. EK~Draconis (HD~129333)}
\author[I. A. Waite et al.]{I.A.~Waite$^{1}$, S.C.~Marsden$^{1}$, B.D.~Carter$^{1}$, P.~Petit$^{2,3}$, S.V.~Jeffers$^{4}$, J. Morin$^{5}$ \and A.A. Vidotto$^{6}$, J.-F.~Donati$^{2,3}$  and the BCool Collaboration  \thanks 
{E-mail:waite@usq.edu.au} \\
${^1}$Computational Engineering and Science Research Centre, University of Southern Queensland, Toowoomba, 4350, Australia    \\
${^2}$Universit\'{e}de Toulouse, UPS-OMP, Institut de Recherche en Astrophysique et Plan\'{e}tologie, Toulouse, France 	\\
${^3}$CNRS, Institut de Recherche en Astrophysique et Plan\'{e}tologie, 14 Avenue Edouard, Belin, F-31400, Toulouse, France \\
${^4}$Institut f\"{u}r Astrophysik, Georg-August-Universit\"{a}t G\"{o}ttingen, Friedrich-Hund-Platz 1, 37077 G\"{o}ttingen, Germany \\
${^5}$LUPM, Universit{\'e} de Montpellier, CNRS, France \\
${^6}$School of Physics, Trinity College Dublin, University of Dublin, Dublin-2, Ireland  \\
}

\begin{document}

\date{}

\pagerange{\pageref{firstpage}--\pageref{lastpage}} \pubyear{2002}

\maketitle

\label{firstpage}

\begin{abstract}
The magnetic fields, activity and dynamos of young solar-type stars can be empirically studied using time-series of spectropolarimetric observations and tomographic imaging techniques such as Doppler imaging and Zeeman Doppler imaging. In this paper we use these techniques to study the young Sun-like star EK~Draconis (SpType: G1.5V, HD~129333) using $\it{ESPaDOnS}$ at the Canada-France-Hawaii Telescope and $\it{NARVAL}$ at the T\'{e}lescope Bernard Lyot. This multi-epoch study runs from late 2006 until early 2012. We measure high levels of chromospheric activity indicating an active, and varying, chromosphere. Surface brightness features were constructed for all available epochs. The 2006/7 and 2008 data show large spot features appearing at intermediate-latitudes. However, the 2012 data indicate a distinctive polar spot. We observe a strong, almost unipolar, azimuthal field during all epochs that is similar to that observed on other Sun-like stars. Using magnetic features, we determined an average equatorial rotational velocity, $\Omega_{eq}$, of $\sim$2.50~$\pm$~0.08 \rdd. High levels of surface differential rotation were measured with an average rotational shear, $\Delta\Omega$, of $\sim$0.27$_{-0.26}^{+0.24}$~\rdd.  During an intensively observed 3-month period from December 2006 until February 2007, the magnetic field went from predominantly toroidal ($\sim$80\%) to a more balanced poloidal-toroidal ($\sim$40-60\%) field.  Although the large-scale magnetic field evolved over the epochs of our observations, no polarity reversals were found in our data. 

\end{abstract}

\begin{keywords}
Line: profiles - stars: activity - stars: individual: HD~129333 - stars: magnetic fields - stars: solar-type - starspots.
\end{keywords}

\section{Introduction}

In Sun-like stars, the generation of the magnetic field is via a dynamo process, with differential rotation being one of the key drivers. The classical dynamo model for the Sun is believed to be operating at the interface between the radiative zone and the convective zone and is known as a $\alpha-\Omega$ or ``shell'' dynamo \citep[e.g.][]{1955ApJ...122..293P,2010LRSP....7....3C}. While the large-scale toroidal magnetic field is understood to be buried in the sub-surface layers of the Sun; it is observed at the surface of a range of rapidly rotating solar-type stars through the presence of strong unipolar surface azimuthal magnetic fields \citep[e.g.][]{2003MNRAS.345.1145D,2004MNRAS.348.1175P,2006MNRAS.370..468M}. One explanation is that these stars distribute the dynamo action closer to the surface of the star \citep[e.g.][]{1989A&A...213..411B,1995A&A...294..155M,2010ApJ...711..424B}. Detailed 3-dimensional magnetohydrodynamic modelling using anelastic spherical harmonic code produces models of azimuthal field wreaths that are similar to the ring-like surface field structures observed on rapidly rotating solar-type stars \citep[e.g.][]{2011ApJ...731...69B,2013ApJ...762...73N}.

Given the key role of differential rotation on magnetic field generation with increasing stellar rotation, long-term measurements of magnetic field topologies and differential rotation for individual stars potentially provide a way to survey the emerging stellar magnetic cycles of young stars, and so to indirectly study the origins of the solar cycle \citep[e.g.][]{2010LRSP....7....3C}. This makes differential rotation and magnetic topologies key tools for the study of the activity of Sun-like stars. These provide a deeper understanding of stellar evolution and in particular the operation and evolution of stellar magnetic fields and the likely impact on any emergent planetary systems \citep[e.g][]{2016ApJ...820L..15D}. 

The focus of this investigation is the young Sun-like star EK~Draconis (HD~129333, HIP~71631). It is an ideal proxy of the infant Sun at a near- zero-age main-sequence (ZAMS) age and as a result, has been the subject of many campaigns using a range of telescopes such as NASA's Hubble Space Telescope \citep[e.g.][]{2010ApJ...723L..38A,2012ApJ...745...25L}, the Far Ultraviolet Spectroscopic Explorer \citep*[e.g.][]{2003ApJ...594..561G}, the Extreme Ultraviolet Explorer \citep*[e.g.][]{1999ApJ...513L..53A}, the R\"{o}ntgensatellit X-ray observatory \citep[e.g.][]{1995A&A...301..201G} and ESA's X-ray Multi-Mirror Mission \citep[e.g.][]{2005A&A...432..671S} to name a few. Additionally, a range of techniques have been utilized such as speckle interferometry \citep{2005A&A...435..215K}, direct imaging \citep{2004ApJ...617.1330M}, photometry (both broadband and Str\"{o}mgren) \citep[e.g.][]{2002A&A...391..659F,2005SerAJ.170..111Z}, Doppler imaging (DI) \citep[e.g.][]{1998A&A...330..685S,2007A&A...472..887J,2016A&A...593A..35R}, and now in this paper we reconstruct its large-scale magnetic field geometry using Zeeman Doppler imaging (ZDI).

The application of ZDI \citep{1989A&A...225..456S,1990SoPh..128..227D,1997A&A...326.1135D,2003MNRAS.345.1145D} enables us to indirectly observe the large-scale surface magnetic field geometry. The mapping of the magnetic fields on EK~Draconis is important as previously only brightness maps have been developed for this young Sun-like star \citep{1998A&A...330..685S,2007A&A...472..887J}. Magnetic imaging will assist with the long-term goal of understanding the early magnetic life of our Sun and its influences on the young emerging planetary systems \citep[e.g.][]{2014ApJ...790L..23D,2016ApJ...820L..15D}. This paper continues a series of papers that investigates the magnetic field topologies of moderately rotating, young Sun-like stars with HD~35296 and HD~29615 the subjects of paper I of this series \citep{2015MNRAS.449....8W} (hereinafter, Paper I). This study of EK~Draconis is part of the BCool\protect\footnote{http://bcool.ast.obs-mip.fr} collaboration investigating the magnetic activity of low-mass stars \citep[e.g.][]{2014MNRAS.444.3517M}.   

\section{Previous studies of EK Draconis}

EK Draconis is a G1.5V star \citep{2001MNRAS.328...45M} and is considered a good proxy for a young Sun \citep[e.g.][]{1994ApJ...428..805D,1998A&A...330..685S,2007A&A...472..887J}. The $\it{HIPPARCOS}$ space mission measured a parallax of 29.30~$\pm$~0.37 mas \citep{2007A&A...474..653V}. \citet{2001MNRAS.328...45M} suggest that EK~Draconis is a member of the Local Association based on its space motion, high levels of activity and strong Li $\textsc {ii}$ with a equivalent width of 189~m\AA.  \citet{2000A&A...355.1087G} estimated its age to be between 30-50 Myr. 

For some time, EK~Draconis was considered an infant sun \citep{1994ApJ...428..805D}. \citet{2007A&A...472..887J} observed the wings of the 866.2~nm Ca \textsc{ii} Infrared Triplet line and concluded that the photosphere is very similar to that of the Sun. \citet{2015A&A...573A..67P} estimated an effective temperature of 5561~K for the primary component while \citet{2007A&A...472..887J} determined the microturbulence to be $\xi_t$ = 1.6~$\pm$~0.1~\kms\ with a metallicity of $[M/H]$ = 0.0~$\pm$~0.05. \citet{1991A&A...248..485D} suggested that EK~Draconis was a binary with a secondary component of mass $\ge$~0.37 $M_{\sun}$. \citet{2004ApJ...617.1330M} used adaptive optics on the 5-m Palomar Telescope to directly image EK~Draconis to confirm the existence of the secondary component. They defined the primary component to be 0.9~$\pm$~0.1 $M_{\sun}$ with the secondary being 0.5~$\pm$~0.1~$M_{\sun}$ in a highly eccentric (e = 0.82~$\pm$~0.03) orbit. Additionally, they speculated on the possibility of EK~Draconis being a triple system but \citet{2005A&A...435..215K} used speckle interferometry to rule out the possibility of a third companion. They also determined the orbital period of this binary system to be $\sim$45~$\pm$~5~years making it unlikely that the two components interact. Thus EK~Draconis is a wide binary whose primary is akin to a single Sun-like star and may be considered an excellent proxy for a young Sun.

EK~Draconis has been the focus of many longitudinal photometric studies, some spanning $\sim$45 years. For example, \citet{1994ApJ...428..805D} found that its activity underwent cyclic variations with an activity cycle of $\sim$12 years. \citet{1995ApJ...438..269B} reported variability, with no apparent periodicity, in the Mount Wilson Ca \textsc{ii} H \& K measurements. \citet*{2005A&A...440..735J} found periodicities in the total spot area on a range of time-scales longer than 45~years, with additional variations with a period of $\sim$10.5 years. \citet{2002A&A...391..659F} have observed a dimming of 0.0057~$\pm$~0.0008~mag~yr$^{-1}$ since 1975, that has even been more pronounced in recent times. \citet{2003A&A...409.1017M} noted the rotational period ranged from 2.551 to 2.886~d which they interpret in terms of surface differential rotation. This all shows EK~Draconis to be an active young Sun.

Doppler imaging maps have been produced by both \citet{1998A&A...330..685S} and \citet{2007A&A...472..887J}. \citet{1998A&A...330..685S} were the first to use DI to map the spot topography on the surface of EK~Draconis. These authors were able to recover high-latitude spots, $\approx$~70-80$\degree$, with a photosphere-to-spot temperature of $\Delta$T~$\approx$~1200~K and mid-latitude spots with $\Delta$T~$\approx$~400~K. \citet{2007A&A...472..887J} produced DI maps that show high latitude spot features, with $\Delta$T~$\sim$500~K. The mean spot latitude of these features varied during the one-year timeframe of their observations, drifting towards the equator at an annual rate of $\approx$~15 to 25$\degree$, depending on the longitude studied.

\begin{table}
\begin{center}
\caption{The parameters used to reconstruct the surface brightness and magnetic field distribution of EK~Draconis. Except otherwise stated, these values were determined by this work.} 
\label{EKDra_parameters}
\begin{tabular}{ll}
\hline
Parameter         			& EK~Draconis                 \\
\hline
Spectral Type 				& G1.5V	$^{a}$				\\
Rotational period			& 2.766~$\pm$~0.002 d 			\\
Inclination Angle 			& 60$\pm$5$\degr$        		\\
\vsini 					& 16.4~$\pm$~0.1 \kmsn  	  	\\
T\subs{phot}				& 5561~K $^{b}$ 			\\
$\Delta$T\subs{phot-spot} 		& 1700~K 				\\
Radial Velocity, v\subs{rad}	 	& -20.28$\pm$~0.04 \kmsn  	  	\\
Stellar radius 				& 0.94~$\pm$~0.07~$R_{\sun}$ 		\\ 
Mass					& 0.95~$\pm$~0.04~$M_{\sun}$ $^{c}$	\\
$\log g$	 				& 4.47~$\pm$~0.08		\\
Stokes $\it{V}$: $\Omega$\subs{eq}	& $\sim$2.50$~\pm$~0.08  \rdd 		\\
Stokes $\it{V}$: $\Delta\Omega$		& $\sim$0.27$_{-0.14}^{+0.12}$ \rdd  	\\
\hline
\end{tabular}
\end{center}
$^{a}$\citet{2001MNRAS.328...45M}; $^{b}$\citet{2015A&A...573A..67P}; $^{c}$based on the theoretical models of \citet*{2000A&A...358..593S}; \\
\end{table}

\section[Observations and Analysis]{Observations and Analysis}

The fundamental parameters used in this study for EK~Draconis are shown in Table \ref{EKDra_parameters}. EK~Draconis was observed using the Canada-France-Hawaii Telescope (CFHT - Mauna Kea, Hawaii) and the T\'{e}lescope Bernard Lyot (TBL - Observatoire du Pic du Midi, France). Observations commenced in November 30, 2006 and the star was periodically revisited until February 9, 2012. A journal of observations is shown in Tables \ref{spectrocopy_log_CFHT} and \ref{spectroscopy_log_TBL}. The data were obtained from POLARBASE\protect\footnote{http://polarbase.irap.omp.eu} \citep{2014PASP..126..469P}.

\subsection{High Resolution Spectropolarimetric Observations from the CFHT and TBL}

High resolution spectropolarimetric data were obtained using ESPaDOnS \citep{2006ASPC..358..362D} at the CFHT and NARVAL \citep{2003EAS.....9..105A}, the twin of ESPaDOnS, at the TBL. Each instrument has a mean pixel resolution of 1.8 \kms\ per pixel. The spectral coverage was from $\sim$ 370 to 1048~nm with a resolution of $\sim$68000. The grating has 79~gr/mm with a 2k~$\times$~4.5k~CCD detector covering 40 orders (orders \#22 to \#61). This extends to the Ca $\textsc{ii}$ H \& K lines and also to the Ca \textsc{ii} Infrared Triplet (IRT) lines. Each instrument consists of one fixed quarter-wave retarder sandwiched between two rotating half-wave retarders and coupled to a Wollaston beamsplitter. Each polarimetric sequence consists of four sub-exposures. After each sub-exposure, the rotating half-wave retarder of the polarimeter was rotated so as to remove instrumental polarization signals from the telescope and the polarimeter. \citet{2012MNRAS.426.1003S} demonstrated the stability of both instruments over a lengthy period of time; hence these are ideal instruments for multi-epoch studies spanning several years. 


\begin{table}
\begin{center}
\caption{The Journal of spectropolarimetric observations of EK~Draconis using the CFHT.} 
\label{spectrocopy_log_CFHT}
\begin{tabular}{lllll}
\hline
UT Date &  UT          & Exp. Time$^{a}$ & Stokes \it{V} & Stokes \it{I}   \\
        &  middle      &    (sec)  & SNR 	& reject.$^{b}$	\\
\hline
2006 Nov 30 & 16:17:34 & 4$\times$60  & 4310  & 1,2,3,4\\
2006 Dec 05 & 16:10:53 & 4$\times$60  & 5384  & 3,4 \\
2006 Dec 06 & 16:12:30 & 4$\times$100 & 8291  & 3,4\\
2006 Dec 07 & 16:09:35 & 4$\times$180 & 10850 & 4\\
2006 Dec 07 & 16:21:37 & 4$\times$60  & 6415  & 1,2,3,4\\
2006 Dec 08 & 16:12:36 & 4$\times$150 & 8254  & 2,3,4\\
2006 Dec 09 & 16:15:40 & 4$\times$120 & 8649  & 2,3,4\\
2006 Dec 10 & 16:12:58 & 4$\times$150 & 9157  & 3,4\\
2006 Dec 11 & 16:12:43 & 4$\times$120 & 8063  & 3,4\\
\hline
\end{tabular}
\end{center}
$^{a}$ Each sequence consists of four sub-exposures with each sub-exposure being 60 s (for example). \\
$^{b}$ The frame number of each Stokes $\it{I}$ sub-exposure that was rejected due to solar contamination, as explained in Sect. \ref{SolarCont} \\
\end{table}


\begin{table}
\begin{center}
\caption{The Journal of spectropolarimetric observations of EK~Draconis using the TBL.} 
\label{spectroscopy_log_TBL}
\begin{tabular}{llllll}
\hline
UT Date &  UT           & Exp. Time$^{a}$  & Stokes $\it{V}$ & Stokes $\it{I}$    \\ 
        &  middle       &    (sec)  & SNR & reject.$^{b}$	\\
\hline
2007 Jan 26 &  6:39:37 & 4$\times$300 & 6446 &  2,3,4   \\  
2007 Jan 27 &  5:49:08 & 4$\times$300 & 7058 &  --    \\
2007 Jan 28 &  5:18:08 & 4$\times$300 & 7603 &  --      \\  
2007 Jan 29 &  5:16:09 & 4$\times$300 & 9244 &  --   \\
2007 Jan 30 &  5:07:46 & 4$\times$300 & 6215 &  --   \\
2007 Feb 01 &  4:44:47 & 4$\times$300 & 6240 &  --   \\
2007 Feb 02 &  5:54:59 & 4$\times$300 & 8026 &  --    \\
2007 Feb 03 &  5:32:38 & 4$\times$300 & 8522 &  --   \\
2007 Feb 04 &  5:49:48 & 4$\times$300 & 9129 &  --   \\ 
2007 Feb 15 &  6:20:12 & 4$\times$300 & 7531 &  3,4  \\
2007 Feb 18 &  4:16:06 & 4$\times$300 & 925  &  4$^{c}$  \\
2007 Feb 19 &  3:10:57 & 4$\times$300 & 7079 &  --  \\
2007 Feb 22 &  2:27:53 & 4$\times$300 & 8469 &  --  \\
2007 Feb 23 &  3:10:56 & 4$\times$300 & 8484 &  --  \\
2007 Feb 28 &  1:59:43 & 4$\times$300 & 1517 &  3,4$^{c}$ \\
\hline
2008 Jan 23 &  6:34:03 & 4$\times$300 &  9772 &  3,4 \\
2008 Jan 24 &  6:27:32 & 4$\times$300 &  7065 &  2,3,4 \\
2008 Jan 26 &  6:07:26 & 4$\times$400 &  10997 & --   \\
2008 Jan 27 &  6:22:54 & 4$\times$300 &  10978 & --  \\
2008 Jan 28 &  6:41:25 & 4$\times$300 &  11937 & --  \\
2008 May 26 & 20:39:17 & 4$\times$300 &  10512 & --   \\
2008 May 26 & 21:08:42 & 4$\times$400 &  11809 & --  \\
2008 May 29 & 21:04:03 & 4$\times$300 &  14546 & --  \\
2008 May 29 & 21:54:48 & 4$\times$550 &  14468 & --  \\
\hline
2009 Jan 3  & 5:53:33 & 4$\times$600 &  17250  & --\\
2009 Jan 4  & 5:24:57 & 4$\times$600 &  14155  & --\\
2009 Jan 5  & 5:22:01 & 4$\times$600 &  15814  & --\\
2009 Jan 11 & 4:51:47 & 4$\times$600 &  12926  & -- \\
\hline
2010 Feb 15 & 5:53:00 & 4$\times$300 & 9510 & \\
\hline
2012 Jan 12 &  6:11:54 & 4$\times$300 & 9959  & --  \\
2012 Jan 13 &  5:37:09 & 4$\times$300 & 9502  & -- \\
2012 Jan 14 &  5:20:28 & 4$\times$300 & 8206  & --  \\
2012 Jan 15 &  5:20:07 & 4$\times$300 & 8149  & --  \\
2012 Jan 16 &  5:01:48 & 4$\times$300 & 7770  & --  \\
2012 Jan 17 &  4:36:18 & 4$\times$300 & 6175  & --  \\
2012 Jan 18 &  4:28:34 & 4$\times$300 & 8276  & --  \\
2012 Jan 23 &  5:59:36 & 4$\times$300 & 8930  & --  \\
2012 Jan 24 &  5:06:37 & 4$\times$300 & 8127  & --  \\
2012 Feb 09 &  1:25:37 & 4$\times$300 & 7797  & --  \\
\hline 
\end{tabular}
\end{center}
$^{a}$ Each sequence consists of four sub-exposures with each sub-exposure being 60 s (for example). \\
$^{b}$ The frame number of each Stokes $\it{I}$ sub-exposure that was rejected due to solar contamination, as explained in Sect. \ref{SolarCont} \\
$^{c}$ In these cases, LSD profiles were rejected due to poor signal-to-noise of the spectrum. \\
\end{table}


\subsection{Spectropolarimetric Analysis}

The initial data reduction was completed using the dedicated pipe-line {\small LibreES{\textsc p}RIT} (\'{E}chelle Spectra Reduction: an Interactive Tool) software package \citep{1997MNRAS.291..658D,2003MNRAS.345.1145D}. Preliminary processing involved removing the bias and using a nightly master flat. Each stellar spectrum was extracted and wavelength calibrated against a Thorium-Argon lamp. After using {\small LibreES{\textsc p}RIT}, Least Squares Deconvolution (LSD) was applied to the reduced spectra. LSD is a multi-line technique that combines several thousand spectral lines into a single line profile with greatly improved signal-to-noise ratio. LSD can be applied to both Stokes $\it{I}$ and $\it{V}$ spectra.
A G2 line mask created from the Kurucz atomic database and ATLAS9 atmospheric models \citep{1993KurCD..13.....K,1993KurCD..18.....K} was used during the LSD process.  In order to correct for the minor instrumental shifts in wavelength space as a result of small atmospheric temperature and pressure variations, each spectrum was shifted to match the Stokes {\it I} LSD profile of the telluric lines contained in the spectra, as was done by \citet{2003MNRAS.345.1145D}, \citet{2006MNRAS.370..468M} and others. Further information on LSD can be found in \citet{1997MNRAS.291..658D} and \citet{2010A&A...524A...5K}.

\subsection{Solar Contamination}
\label{SolarCont}
There was some severe solar contamination in the red wing and core of the Stokes ${\it I}$ LSD profile for the CFHT data (most nights) and some contamination in the TBL data due to the observations being taken close to sunrise. To determine the influence of sunrise on the contamination, the first Stokes $\it{I}$ profile of the sequence of four sub-exposures was used as the template with successive profiles individually subtracted, producing a difference profile. If the deviation of this difference profile exceeds the average noise level of the first LSD profile in the areas of concern, this Stokes $\it{I}$ profile was excluded from the brightness mapping process. Those profiles excluded are listed in Tables \ref{spectrocopy_log_CFHT} and \ref{spectroscopy_log_TBL}. All Stokes $\it{I}$ profiles taken on the November 30, 2006 were rejected as the first profile was severely affected as were the second sequence of exposures taken on December 07, 2006. We searched for evidence of contamination in the Stokes $\it{V}$ profiles by examining the null profile. The null profile is used as a check for spurious magnetic signatures and is generated by ``pair-processing" sub-exposures corresponding to the identical positions of the half-wave Fresnel Rhomb of the polarimeter during each sequence of four sub-exposures \citep[see][]{1997MNRAS.291..658D}. There appeared no contamination in the Stokes $\it{V}$ data exceeding the typical noise level of the data. 

\section{Chromospheric Activity}
\label{chromo}

Chromospheric activity was measured using the Ca $\textsc{ii}$ H \& K, Ca $\textsc{ii}$ Infrared Triplet (IRT) and H$\alpha$ spectral lines. The two Ca $\textsc{ii}$ H \& K absorption lines are the most widely used optical indicators of chromospheric activity as their source functions are collisionally controlled and hence they are very sensitive to electron density and temperature. This leads to emission in the core of these lines. The  Ca $\textsc{ii}$ IRT lines share the upper levels of the H \& K transitions and are formed in the lower chromosphere \citep[e.g.][]{2004LNEA....1..119M}. The H$\alpha$ spectral line is also collisionally filled in as a result of higher temperatures and is formed in the middle of the chromosphere and is often associated with plages and prominences \citep[e.g.][]{1993MNRAS.262....1T,2004LNEA....1..119M}. 

The Ca $\textsc{ii}$ H \& K index for EK~Draconis was determined using the procedure as explained in \citet{2004ApJS..152..261W}. This was converted to match the Mount Wilson S-values \citep{1991ApJS...76..383D} using: 

\begin{equation}
	\label{MountWilson}	
	\ \textrm{S-index} = \frac{C_1 F_H - C_2 F_K}{C_3 F{_{V_{HK}}} + C_4 F{_{R_{HK}}}}-C_{5},  
\end{equation}
where \textit{F$_{H}$} and \textit{F$_{K}$} was the flux determined in the line cores, centred on 393.3663 and 396.8469~nm respectively, from the two triangular bandpasses with a full-width at half maximum (FWHM) of 0.218~nm. Two 2~nm-wide rectangular bandpasses \textit{F$_{R_{HK}}$} and \textit{F$_{V_{HK}}$}, centred on 400.107 and 390.107~nm respectively, were used for the continuum flux in the red and blue sides of the H and K lines. The transformation coefficients used to calibrate these data to the Mt Wilson Survey were determined by \citet{2014MNRAS.444.3517M}. The coefficients for equation \ref{MountWilson} are listed in Table \ref{MW_coeff}.

\begin{table}
\begin{center}
\caption{The coefficients, listed in equation \ref{MountWilson}, as determined by \citet{2014MNRAS.444.3517M}.}
\begin{tabular}{lll}
\hline
Coefficient & ESPaDOnS & NARVAL \\
\hline
C$_{1}$ & 7.999 & 12.873 	\\
C$_{2}$ & -3.904 & 2.502 	\\
C$_{3}$ & 1.150 & 8.877		\\
C$_{4}$ & 1.289 & 4.271   	\\
C$_{5}$ & -0.069 & 1.183 $\times$ 10$^{-3}$ 	\\
\hline
\label{MW_coeff}
\end{tabular}
\end{center}
\end{table}

Two further activity indices were used; the H$\alpha$ spectral line and the Ca \textsc {ii} IRT lines. The H$\alpha$-index was determined using 

\begin{equation}
	\label{Halpha}
		\ \textrm{N$_{H\alpha}$-index} = \frac{F_{H\alpha}}{F_V+F_R},
\end{equation}
where \textit{F$_{H\alpha}$} is the flux determined in the line core, centred on 656.285~nm, using a triangular bandpass with a FWHM of 0.36~nm. Two 0.22~nm-wide rectangular bandpasses \textit{F$_{V}$} and \textit{F$_{R}$}, centred on 655.885 and 656.730~nm respectively, were used for the continuum flux in the blue and red sides of the H$\alpha$ line \citep{2002AJ....123.3356G}. The CaIRT-index was determined using

\begin{equation}
	\label{CaIRT}
		\ \textrm{N$_{CaIRT}$-index} = \frac{\sum F_{IRT}}{F_V+F_R},
\end{equation}
where $\sum F_{IRT}$ is the total flux determined in the line cores of the three spectral lines, 849.8023, 854.2091 and 866.2141~nm, using triangular bandpasses with a FWHM of 0.2~nm. Two 0.5~nm-wide rectangular bandpasses \textit{F$_V$} and \textit{F$_R$}, centred on 847.58 and 870.49~nm respectively, were used for the continuum flux in the blue and red sides of the Ca \textsc {ii} IRT spectral lines \citep{2013LNP...857..231P}.

\begin{figure*}
\begin{center}
\begin{subfigure}[]{0.49\textwidth}
	\includegraphics[width=\textwidth]{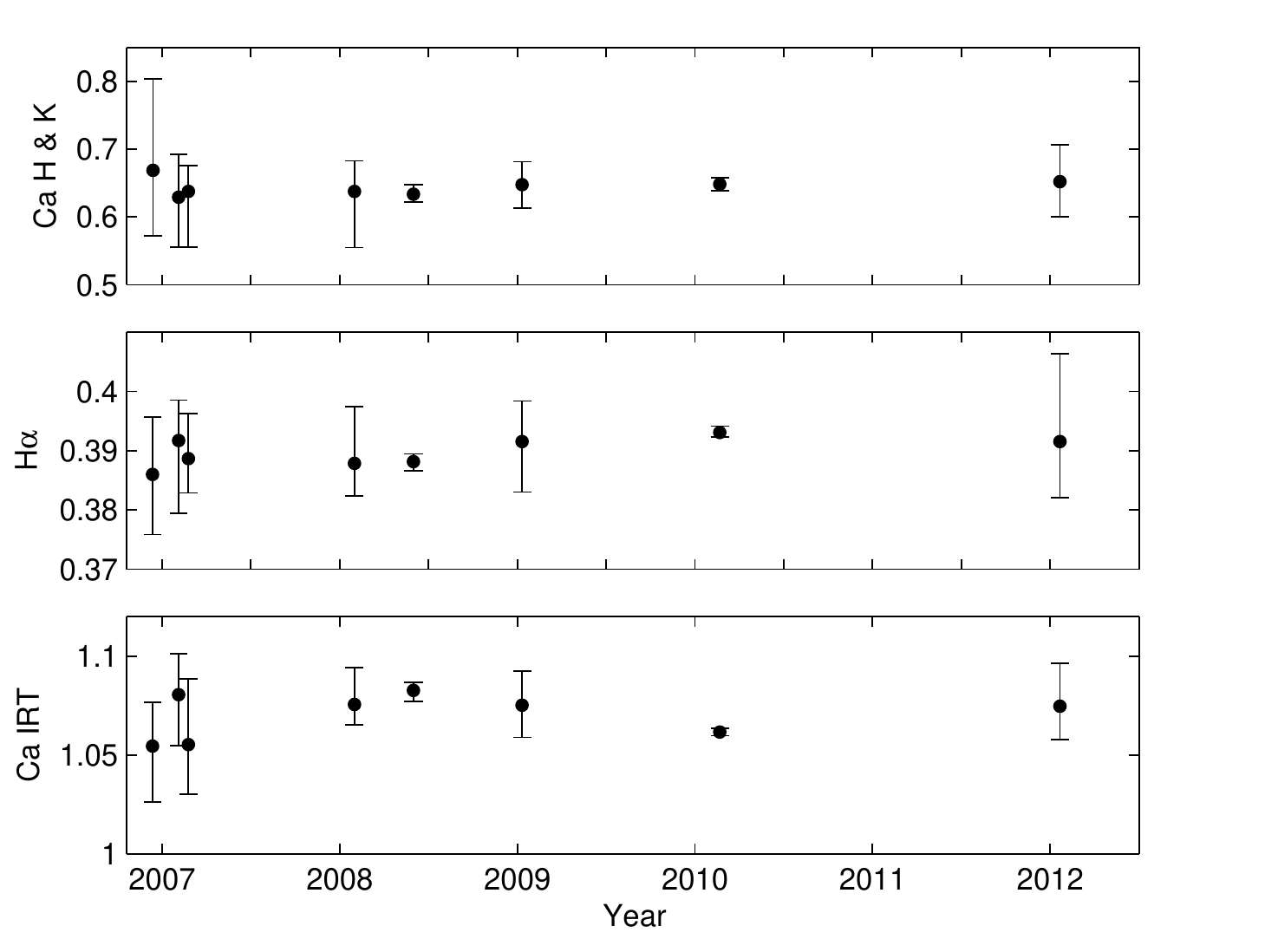}
	\caption{Long-term activity variations 2006 -- 2012}
	\label{fig:Chromospheric_all}
\end{subfigure}
\begin{subfigure}[]{0.49\textwidth}
	\includegraphics[width=\textwidth]{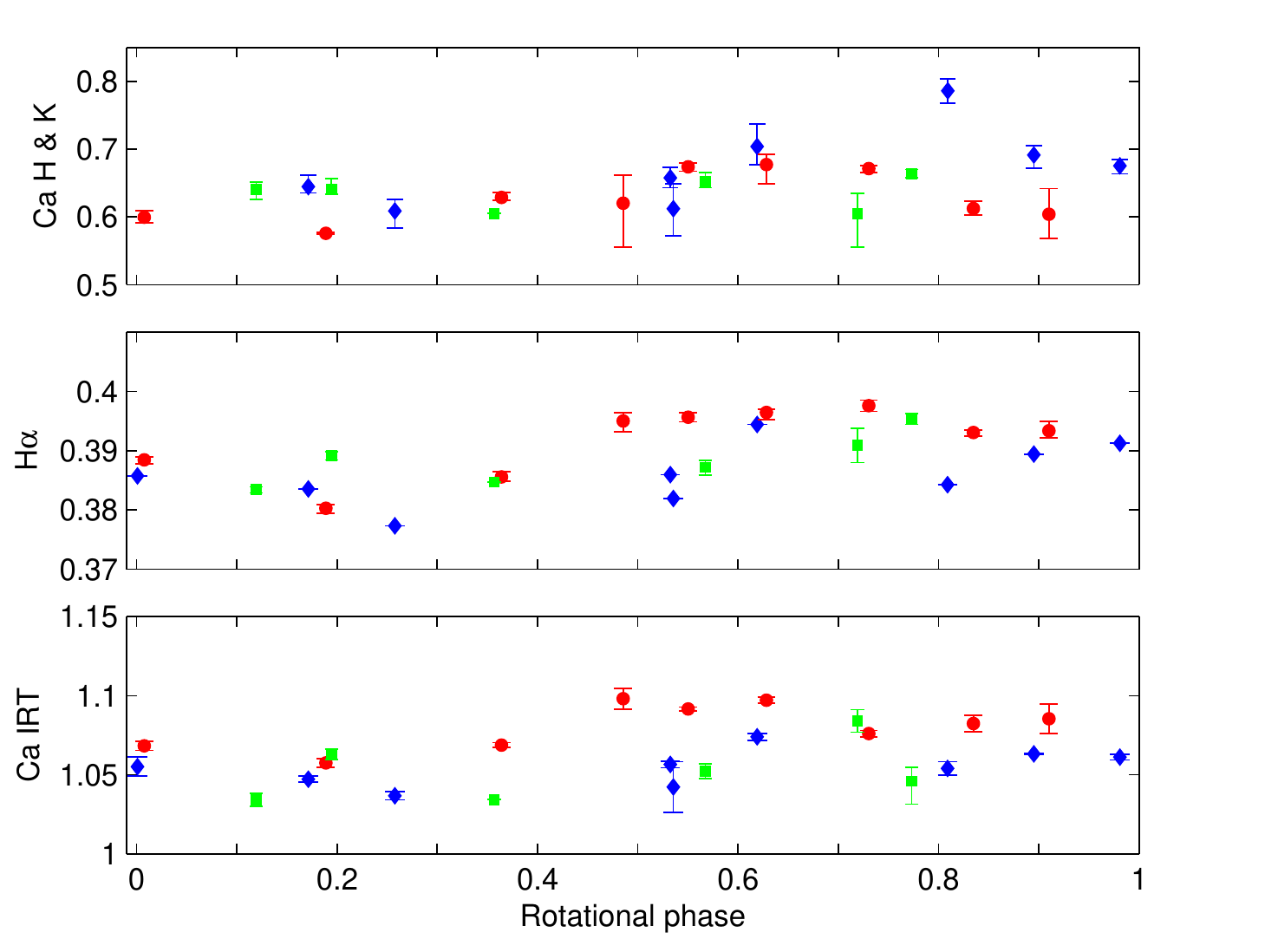}
	\caption{Short-term activity variations 2006 -- 2007}
	\label{fig:Chromospheric_3months}
\end{subfigure}
\caption{The left series of panels shows the long term variations of the activity indices from December 2006 until January 2012. Each data point was an average of all the respective indices for a particular dataset while the error bars provide the range of the activity indices. The right series of panels shows the variation in the chromospheric activity over the intensely observed 2006/2007 observing run. The data were phased to the ephemeris using equation \ref{ephemeris} Diamonds represent CFHT, circles TBL January-early February observations and squares TBL late February 2007. The error bars indicate the range of the measurement during four sub-exposures.}
\label{fig:AllChromo}
\end{center}
\end{figure*}

Fig. \ref{fig:Chromospheric_all} shows the variation in the Ca $\textsc{ii}$ H \& K lines, H$\alpha$ and Ca $\textsc{ii}$ IRT spectral lines during the five year interval that these data cover. Using the values from \citet*{2009A&A...493.1099S}, $\log R^\prime_{HK}$ was determined as -4.08~$\pm$~0.043. Table \ref{activity_indices} show the average S-index, N$_{H\alpha}$ and N$_{CaIRT}$ indices for the respective datasets. Each value is the average over the observing run with the range showing the variation during this time, most likely as a result of modulation due to the rotation of the star. Additionally, the number of exposures included varies, as listed in Tables \ref{spectrocopy_log_CFHT} and \ref{spectroscopy_log_TBL}. The phases, $\phi$, of the observations were calculated using the ephemeris in equation \ref{ephemeris} 
\begin{equation}
	\label{ephemeris}
	\phi = 2454070.17498 + 2.766E,
\end{equation}
where E is the epoch of each observation.

\begin{table}
\begin{center}
\caption{Average activity indices found for EK~Draconis. Also listed is the range of the respective index for the observing season. Additionally, the number of exposures included varies, as listed in Tables \ref{spectrocopy_log_CFHT} and \ref{spectroscopy_log_TBL}.}
\label{activity_indices}
\begin{tabular}{llll}
\hline
Date	&  S-index$^{a}$ & N$_{H\alpha}$ & N$_{CaIRT}$ \\
\hline
Dec 2006 & 0.669~$\pm$~0.116	&	0.386~$\pm$~0.010	&	1.054~$\pm$~0.025	\\
Jan 2007 & 0.629~$\pm$~0.068	&	0.392~$\pm$~0.010	&	1.081~$\pm$~0.023	\\
Feb 2007 & 0.620~$\pm$~0.199	&	0.389~$\pm$~0.007	&	1.055~$\pm$~0.029	\\
Jan 2008 & 0.637~$\pm$~0.064	&	0.388~$\pm$~0.007	&	1.076~$\pm$~0.015	\\
May 2008 & 0.633~$\pm$~0.013	&	0.388~$\pm$~0.001	&	1.083~$\pm$~0.005	\\
Jan 2009 & 0.648~$\pm$~0.034	&	0.392~$\pm$~0.008	&	1.075~$\pm$~0.017	\\
Feb 2010 & 0.648~$\pm$~0.010	&	0.393~$\pm$~0.001	&	1.062~$\pm$~0.002	\\
Jan 2012 & 0.652~$\pm$~0.053	&	0.392~$\pm$~0.012	&	1.075~$\pm$~0.019	\\
\hline
\end{tabular}
\end{center}
${^a}$The resulting N$_{Ca \textsc{ii}HK}$-index was converted to match the Mount Wilson S-values \citep{1991ApJS...76..383D} using the transformation shown in equation \ref{MountWilson}. \\
\end{table}

Focusing on the three month period extending from December 2006 until February, 2007, the indices show variations based on the rotation of EK~Draconis. This is shown in Fig. \ref{fig:Chromospheric_3months}. There are variations due to spot evolution; for example, there is enhanced activity in the January 2007 observing season when compared with the December 2006 season, especially the N$_{CaIRT}$ index (bottom panel) at phase $\sim$0.5. 

The average S-index from the Mount Wilson sample was 0.4714~$\pm$~0.0156 and $\log R^\prime_{HK}$ was -4.245~$\pm$~0.088 \citep{1991ApJS...76..383D}. Between 1984 to mid-2003, \citet{2007ApJS..171..260L} measured an average S-index of 0.5475 and $\log R^\prime_{HK}$ of -4.148. At the same time, they observed a decrease in the brightness of the star using Str\"{o}mgren b, y photometry. From late 2006 until early 2012, our project measured an average S-index of 0.644~$\pm$~0.06, and the $\log R^\prime_{HK}$ of -4.08~$\pm$~0.043. This is substantially higher than that of the Mount Wilson Ca H~\&~K project but closer to that of \citet{2007ApJS..171..260L}. We infer that the activity on EK~Draconis has increased since 1991 matching the reduction in brightness observed by \citet{2002A&A...391..659F} and \citet{2007ApJS..171..260L}. EK~Draconis continues to exhibit both short and long-term variations, as shown in Fig. \ref{fig:AllChromo}, and alluded to by \citet{1995ApJ...438..269B} during the Mount Wilson Ca $\textsc{ii}$ H \& K project. However, we cannot confirm the 12-year cyclic period reported by \citet{1994ApJ...428..805D} due to the sparseness of observations during our five year time-frame.

\section{Image Reconstruction}
\label{DI_ZDI}
\subsection {Doppler Imaging: Brightness maps}
\label{DI}

Surface brightness images of this moderately rotating star were produced for five epochs ranging from December 2006 until January 2012. Each of these maps, assuming solid-body rotation, were generated through the inversion of time series Stokes $\it{I}$ LSD profiles. The imaging code used was that of \citet{1991A&A...250..463B} and \citet{1997A&A...326.1135D}. This inversion process is an ill-posed problem where an infinite number of solutions is possible. Therefore, this code implements the \citet{1984MNRAS.211..111S} maximum-entropy optimization that produces an image with the minimum amount of information. Synthetic Gaussian profiles are often used to produce the initial model \citep{1995MNRAS.273....1U}, although these are not as effective for slow and moderate rotators. Hence the initial modelled profiles of the photosphere and spots were generated using the local line profiles from slowly rotating G2 and K5 stars respectively. This allowed for more effective fitting of the wings of the LSD profile. 

The imaging code was used to establish the values of a number of basic parameters, including the star's projected rotational velocity (\vsini), inclination angle, continuum level and radial velocity. A grid of plausible values was generated and full DI maps produced. The key parameters that were systematically changed were the continuum level (in 0.0001 steps), \vsini\ and v$_{rad}$ (both in 0.1 \kms\ increments) and inclination angle ($\pm$ 5$\degree$). The average LSD profile was then compared with the average modelled LSD profile and deviations between the two were measured across the full profile. By minimizing the deviations at each point on the profile, the best set of parameters were identified. These parameters are listed in Table \ref{EKDra_parameters}.

The imaging code assumes a two-temperature model; one being the quiet photosphere while the other is the temperature of the spot. The estimate of the effective temperature of the photosphere is usually derived from line-depth ratio measurements \citep{1994PASP..106.1248G}, photometric colours (V-I or similar) \citep[e.g.][]{1998A&A...333..231B} or more recently, principal component analysis  \citep{2015A&A...573A..67P}. However, the determination of the temperature of star-spots is more problematic. All spots are assumed to be at the same temperature; hence penumbral influences are neglected. This is necessary to overcome the limitations of rotational blurring and finite signal-to-noise factors \citep{1992LNP...397...33C}. There has been some debate in the literature regarding the temperature of the spots occurring on EK~Draconis. \citet{2007A&A...472..887J} used DI to determine that the spots were only 500~K cooler than the surrounding photosphere. \citet{1998A&A...330..685S} also used DI and found spot temperatures that were 400 to 1200~K cooler than the surrounding photosphere. \citet{1994IBVS.4110....1S} also noted that the spot temperature was approximately 460~K less than the surrounding photosphere using photometric lightcurves. Using the most recent photospheric temperature estimate of \citet{2015A&A...573A..67P} of 5561~K, we started with a relatively warm starspot temperature ($T_{spot}$~=~5161~K) and systematically increased the difference between the spot temperature and the surrounding photosphere ($\Delta$T) to determine the best fit of the modelled data, in a minimum $\chi^2_r$ sense. We found that $\Delta$T~=~1700~K provided the best fits to the data. Any further reduction in the spot temperature made little difference to the resulting fits. This is consistent with the spot temperatures expected of early G-type stars, and supports the relationship found by \citet{2005LRSP....2....8B} (Fig.~7 on page 27 of that work). Further support comes from the work of \citet{2004AJ....128.1802O} who used molecular band modelling, in particular, the TiO Bands at 705.5 and 886.0~nm. They found that the spot temperatures on EK~Draconis were $\sim$ 3800~K; closer to the typical minimum Sunspot umbral temperatures \citep[e.g.][]{2003SoPh..215...87P}. Still, spot temperatures are extremely difficult to estimate based solely on Doppler imaging in the absence additional information such as the inclusion of simultaneous photometry \citep[e.g.][]{2011MNRAS.413.1949W} or more indepth analysis \citep[e.g.][]{2009A&ARv..17..251S}.

\begin{figure}
\begin{center}
\includegraphics[trim=0.4cm 0cm 6cm 0cm, scale=1.15, angle=0]{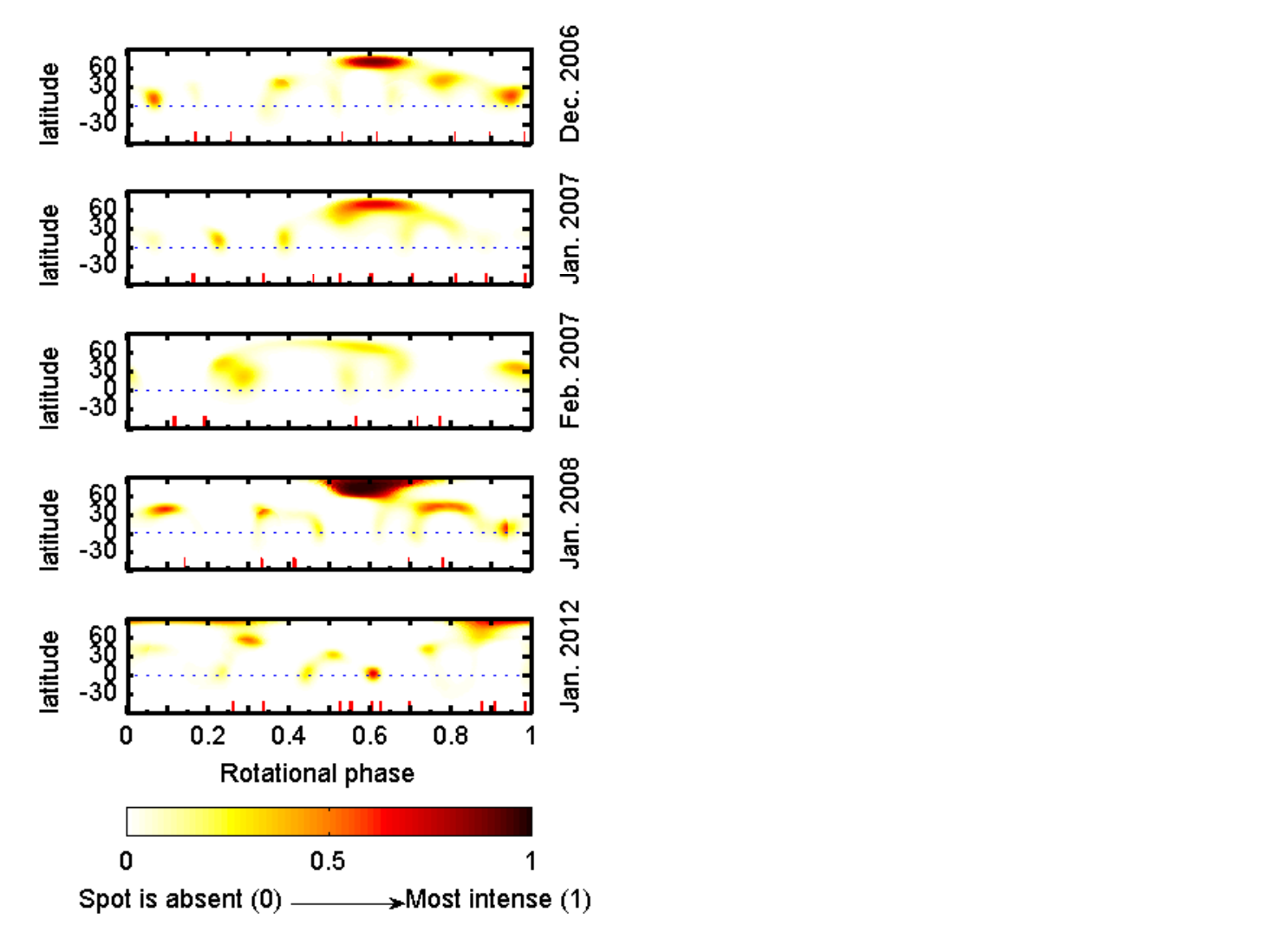}
\caption{Maps of EK~Draconis for data from December 2006 until January 2012. Solid-body rotation was assumed. The larger tick marks (in red) show the phase each observation was taken. The spot filling factors are listed in Table \ref{EKDra_Mapping_Table}. The colour scale at the bottom of the maps indicate the gradation of spottedness with 0 (absent) to 1 (most intense). The associated fits between the modelled LSD profiles that were used to generate these maps and stellar LSD profiles are shown in Fig. \ref{EKDra_StokesI}.}
\label{StokesI_SB} 
\end{center}
\end{figure}

There has been a range of rotational periods found in the literature for EK~Draconis. Many authors use a period of $\sim$2.6 to 2.8~d. \citet{2005A&A...435..215K} used radial velocity variations in their speckle observations to infer a period of 2.767~$\pm$~0.005~d. \citet{2003A&A...409.1017M} used long-term photometry to investigate surface differential rotation and found variations in the period. They found two rotational periods in several seasons, both showing monotonical decreases along with each starspot cycle with solar-like behaviour. \citet{1995A&A...301..201G} reported a period of 2.6$^{+0.4}_{-0.3}$~d in X-rays and a 2.75~$\pm$~0.05~d period in the optical spectrum. They also observed periodicity in the X-ray light curve and, using auto-correlation analysis, found a ``broad secondary maximum around P = 2.53~d''. Assuming solid-body rotation and using the inclination angle of 60$\degree$ \citep{1998A&A...330..685S,2007A&A...472..887J} and a \vsinis of 16.4 \kms (as determined by this work), two separate maps were produced; one for December 2006 and the other for the January 2007. Using the ephemeris in equation \ref{ephemeris}, the rotational period was systematically modified so that the location of the predominant high-latitude spot was at the same phase in both maps. This is shown in the top two panels of Fig. \ref{StokesI_SB}.\footnote{When attempting to model the combined 2006/2007 dataset, the minimum reduced $\chi^2_r$ value was 1.9. This is higher than the typical $\chi^2_r$ values achieved \citep[e.g.][]{2004MNRAS.351..826P,2011MNRAS.413.1922M}. Thus the 3-month long 2006-2007 dataset was split into three separate maps: the CFHT data from November 30, to December 11, 2006, the TBL data from January 25 to February 4, 2007, while the third dataset was taken from February 15 to February 28, 2007.} The rotational period was found to be 2.766~$\pm$~0.002 days; consistent with that of \citet{2005A&A...435..215K}. However, this period determination did not take into consideration differential rotation or spot evolution. Therefore this may be considered the period at ~70$\degree$ latitude and is more analogous to the photometric period obtained using broad-band photometry. Further data were also obtained in 2008 and in 2012. The solid-body rotation maps are shown in Fig. \ref{StokesI_SB}.

Figs. \ref{EKDra_StokesI} show the residual fits to the data. These are normalized profiles as each stellar LSD profile was subtracted from an average stellar LSD profile; likewise, each of the modelled profiles were subtracted from the average modelled profile. Each normalized stellar profile was compared with the associated modelled normalized profile. This has the advantage of accentuating the deviations between the stellar and modelled fits. The $\chi^2_r$ values for each map are listed in Table \ref{EKDra_Mapping_Table}. This modelling assumed solid-body rotation, so small deviations observed between the normalized stellar profiles and the modelled profiles are taken to be as a result of differential rotation and spot evolution. Differential rotation is discussed in Sect. \ref{SDR}.

\begin{figure*}
\begin{center}
\includegraphics[trim=1.0cm 0cm 0cm 0cm, scale=1.30, angle=0]{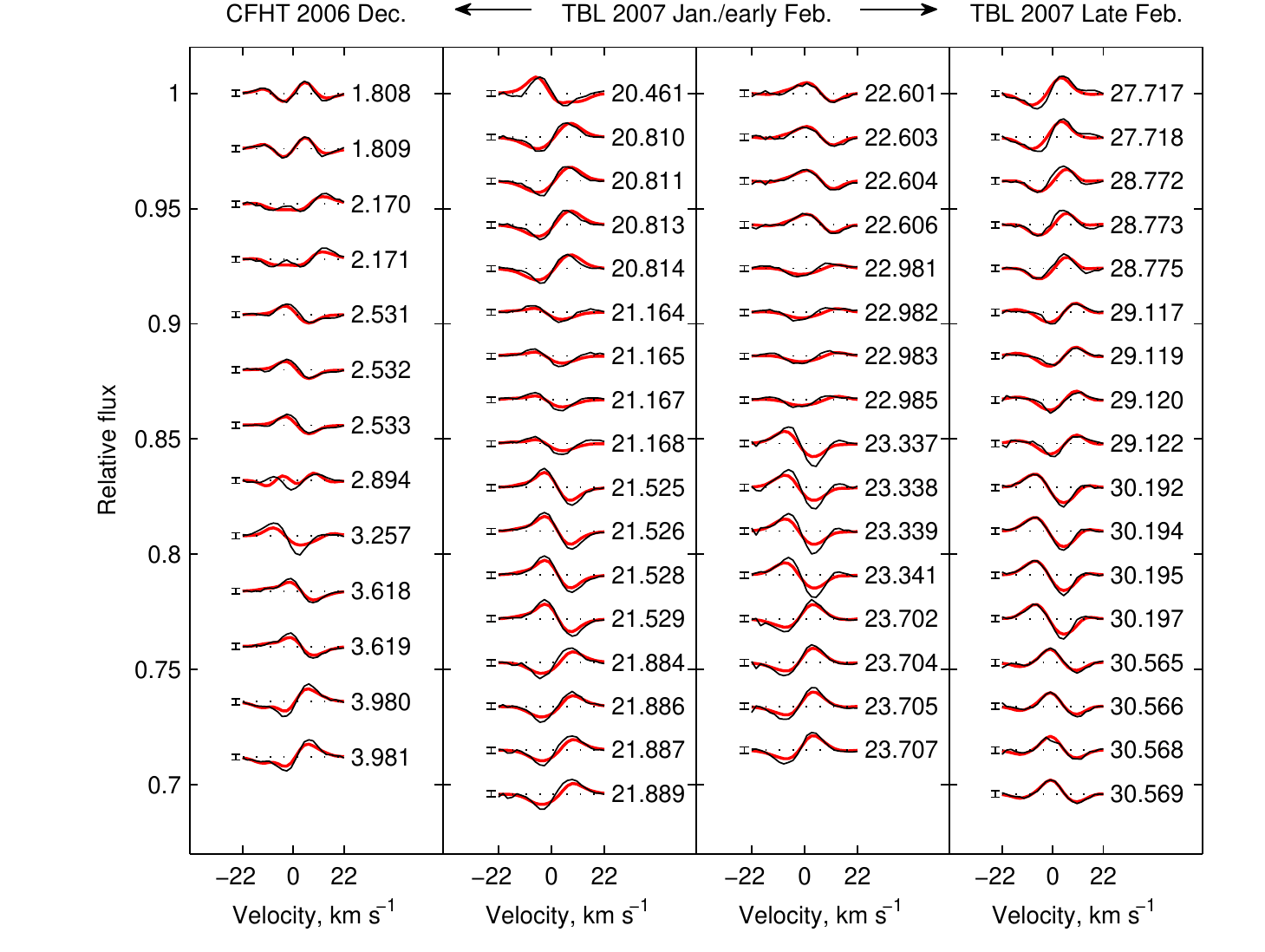}
\caption{The residual maximum-entropy fits to the Stokes $\it{I}$ LSD profiles for EK~Draconis during the December 04, 2006 to February 27 2007 observing run. Differential rotation was not incorporated into the mapping process. These are normalized profiles with the black line showing the normalized raw profile while the red line shows the normalized modelled profile. The error bars on the left indicate the average error, 1-$\sigma$. The numbers on the right represent the phase of each observation, based on the ephemeris in equation \ref{ephemeris}.}
\end{center}
\end{figure*}
\addtocounter{figure}{-1}
\begin{figure*}
\begin{center}
\includegraphics[trim=1.0cm 0cm 0cm 0cm, scale=1.20, angle=0]{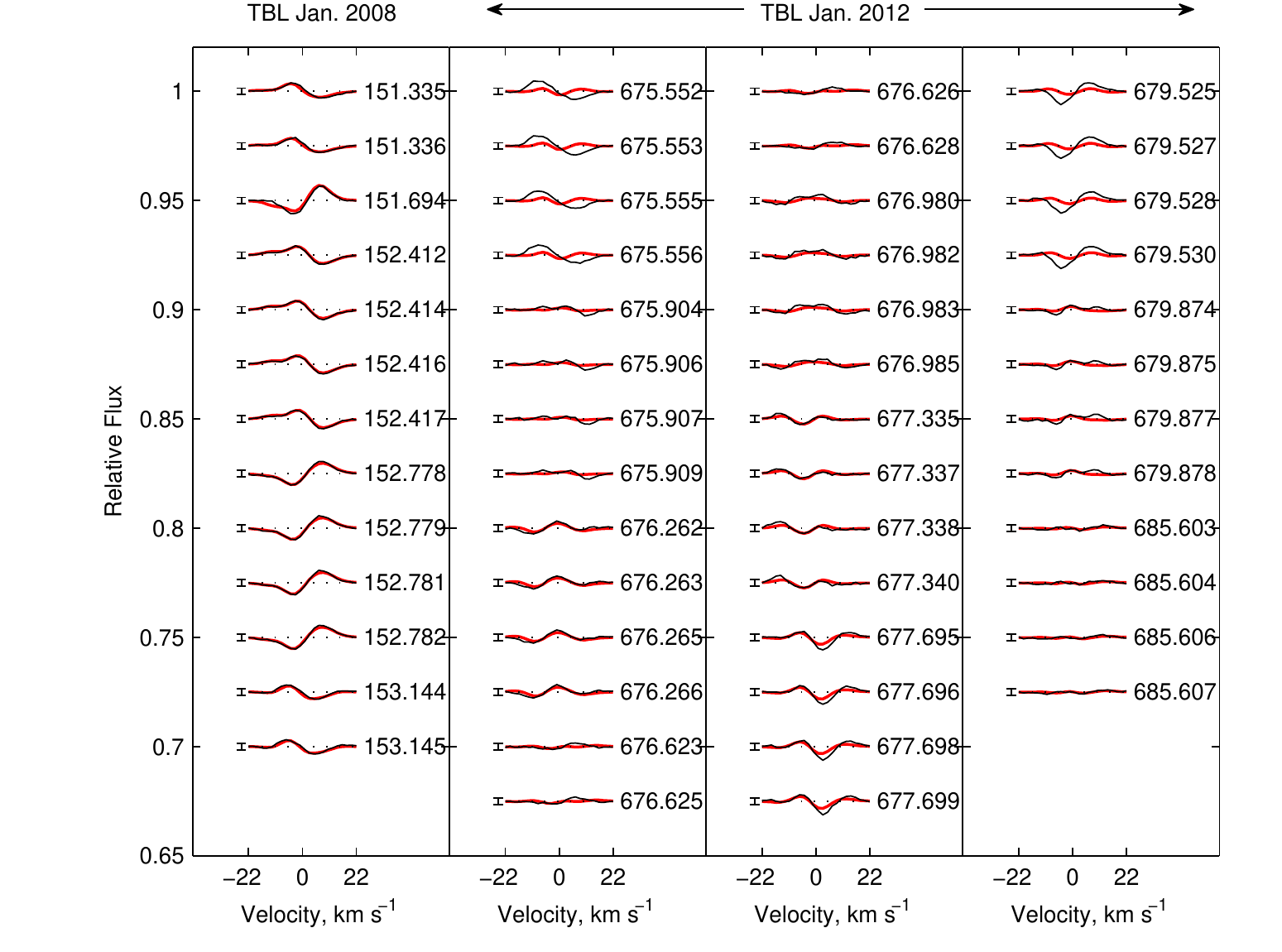}
\caption{(continued) The residual maximum-entropy fits to the Stokes $\it{I}$ LSD profiles for EK~Draconis during the 2008 (left most column) and 2012 observing runs. Differential rotation was not incorporated into the mapping process. These are normalized profiles with the black line showing the normalized stellar profile while the red line shows the normalized modelled profile. The error bars on the left indicate the average error, 1-$\sigma$. The numbers on the right represent the phase of each observation, based on the ephemeris in equation \ref{ephemeris}.} 
\label{EKDra_StokesI} 
\end{center}
\end{figure*}

\begin{table*}
\begin{center}
\caption{The mapping parameters used for EK~Draconis. Column 2 shows the timespan over which the data were taken. Column 3 shows the number of epochs, $\phi_{I}$, (number of observations) used in the mapping process; Columns 4 and 5 lists the $\chi^2_r$ achieved and the spot coverage used in producing the brightness maps from the Stokes $\it{I}$ profiles (see Fig. \ref{StokesI_SB}). The brightness maps assume solid-body rotation. Columns 6 to 10 are the number of epochs, $\phi_{V}$, $\chi^2_r$, average magnetic field strength $<|$B$|>$, $\Omega_{eq}$ and $\Delta\Omega$ used in producing the magnetic maps from the Stokes $\it{V}$ profiles (see Fig. \ref{StokesV_maps}). The final column is the laptime$^{a}$ which is the time the equatorial regions need to lap the pole. Variations (errors) are based on the systematic recalculation of the models based upon varying stellar parameters: $\Omega_{eq}$, $\Delta\Omega$, \vsini, global magnetic field and inclination.}
\label{EKDra_Mapping_Table}
\begin{tabular}{lcccccccccc}
\hline
Year & Timespan & \multicolumn{3}{c}{Stokes$\it{I}$} &  \multicolumn{6}{c}{Stokes$\it{V}$}  \\
\cmidrule(l){3-5} 
\cmidrule(l){6-11}
     & d        & $\phi_{I}$ &  $\chi^2_r$          & spot coverage & 	$\phi_{V}$ & $\chi^2_r$	      & $<|$B$|>$ 	(G)	& $\Omega_{eq}$ (\rdd)		 & $\Delta\Omega$ (\rdd)  & laptime$^{a}$ (d)       \\
\hline	
Dec 2006 & 7 & 7(13)  & 0.70 & 3.7 \% & 8  & 1.35 & 91.8 & 2.42$\pm$0.1 & 0.19$\pm$0.18 & $\sim$33	\\
Jan 2007 & 9 & 9(33)  & 1.00 & 3.3 \% & 9 & 1.40 & 78.3 & 2.52$\pm$0.05 & 0.38$\pm$0.13 & $\sim$16\\
Feb 2007 & 5 & 5(17)  & 0.45 & 3.9 \% & 5 & 1.25 & 57.7 & 2.51$\pm$0.06 & 0.13$_{-0.04}^{+0.2}$ & $\sim$48	\\
Jan 2008 & 5  & 5(15)  & 0.25 & 4.6 \% & 5 & 1.3  & 60.7 & 2.57$\pm$0.03 & 0.39$\pm$0.1  & $\sim$16 \\
Jan 2012 & 28 & 10(40) & 0.70 & 2.6 \% & 10 & 2.0 & 92.3 & 2.44$\pm$0.02 & 0.25$\pm$0.06 & $\sim$25 \\			
\hline	
\multicolumn{11}{|l|}{$^{a}$ The laptime is calculated using \(\frac{2\pi}{\Delta\Omega}\).} \\
\end{tabular}
\end{center}
\end{table*}

\subsection{Zeeman Doppler imaging: Magnetic Mapping}
\label{ZDI}

The magnetic topology was reconstructed using ZDI. The modelling strategy of \citet{1997A&A...326.1135D} was used to construct the radial, azimuthal and meridional fields. The mapping procedure uses the spherical harmonic expansions of the surface magnetic field, as implemented by \citet{2006MNRAS.370..629D}. The maximum spherical harmonic expansion $\ell_{max}$ = 12 was selected as any further increase in $\ell$ did not produce any significant difference in the magnitude and topology of the magnetic field recovered. Differential rotation was measured on all datasets, using the technique explained in Sect.~\ref{SDR}.

\begin{figure*}
\begin{center}
\includegraphics[scale=1.2, angle=0]{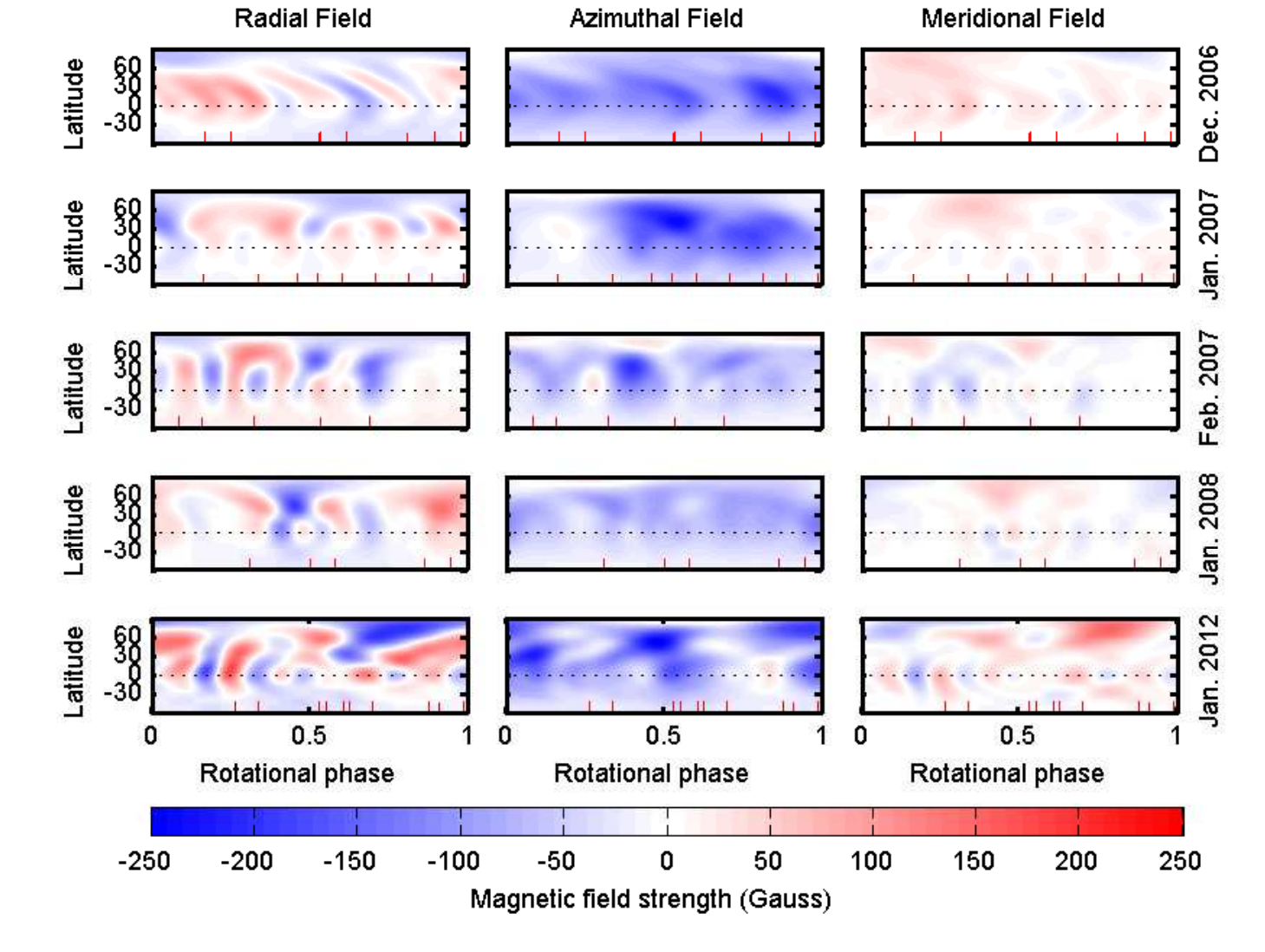}
\caption{Maps of EK~Draconis for the CFHT/TBL from December 2006 until January 2012 data, with the differential rotation parameters incorporated into the imaging process. The extended tick lines (red) along the phase axes indicate the phase of the observation. The colour scale, on the bottom of the maps, is the magnetic field intensity in Gauss. The associated fits between the modelled data that was used to generate these maps and the actual LSD profiles are shown in Fig. \ref{EKDra_StokesV_fits_all}. The LSD profile from 18 February, 2007 was excluded from the mapping process due to poor signal-to-noise.}
\label{StokesV_maps} 
\end{center}
\end{figure*}

\begin{figure*}
\begin{center}
\includegraphics[trim=1.0cm 0cm 0cm 0cm, clip=true, scale=0.85, angle=0]{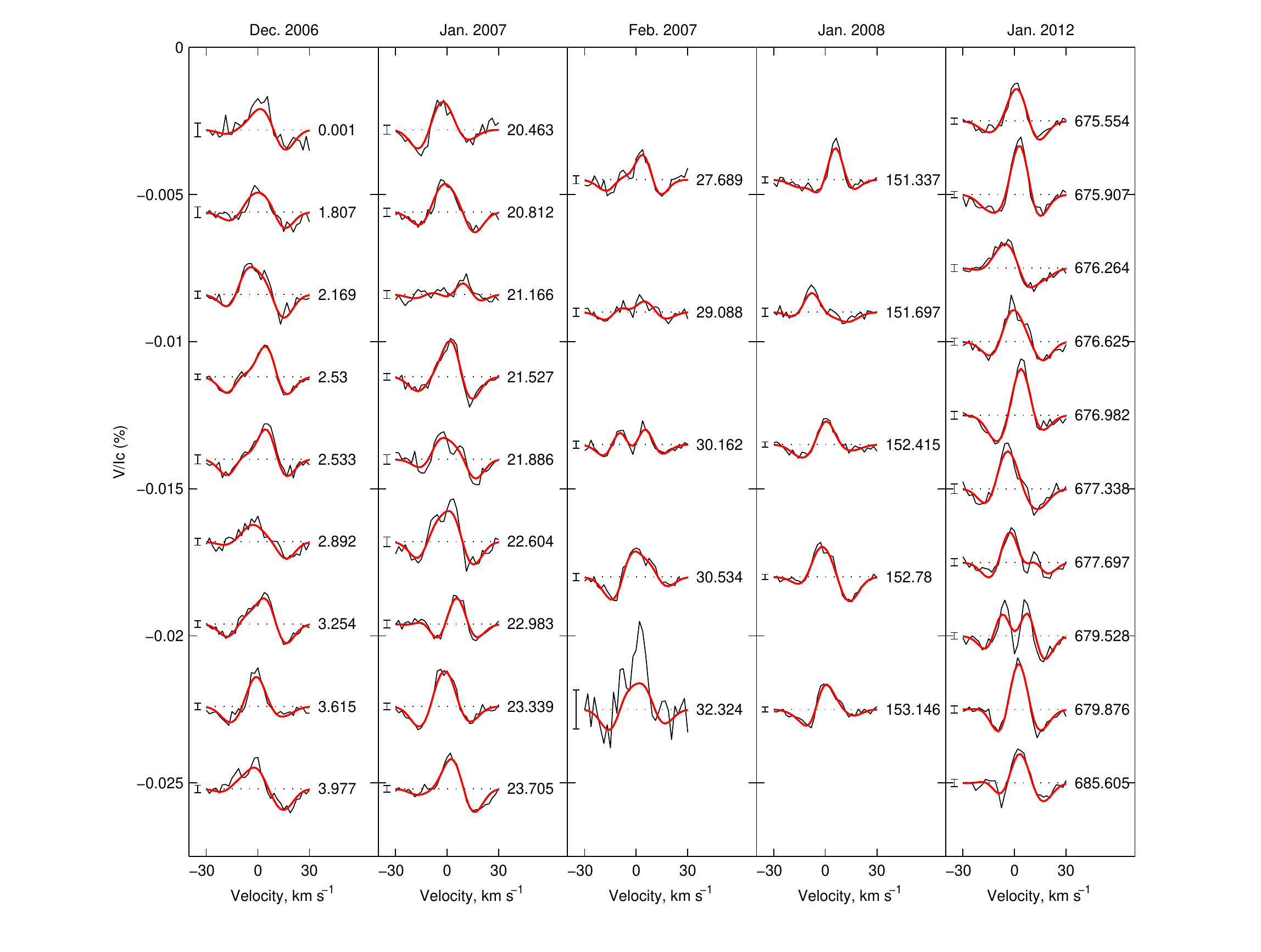}
\caption{The maximum-entropy fits to the Stokes $\it{V}$ LSD profiles for EK~Draconis with surface differential rotation incorporated into the analysis. The thick (red) lines represent the modelled lines produced by the Zeeman Doppler imaging process whereas the thin (black) lines represent the actual observed LSD profiles. Each successive profile has been shifted down by 0.002, or 0.003 when there were fewer profiles, for graphical purposes. The error bar on the left of each profile is plotted to 1-$\sigma$. The rotational phases are indicated to the right of each profile based on the ephemeris in equation \ref{ephemeris}. Each column represents an observation run. The LSD profile from 18 February, 2007 was excluded from the mapping process due to poor signal-to-noise.}
\label{EKDra_StokesV_fits_all}
\end{center}
\end{figure*}

The magnetic maps for EK~Draconis are shown in Fig. \ref{StokesV_maps} while the associated fits between the modelled data and the actual LSD profiles are shown in Fig. \ref{EKDra_StokesV_fits_all}. The $\chi^2_r$ values of the magnetic models are listed in Table \ref{EKDra_Mapping_Table}. The magnetic maps produced for EK~Draconis show complex, and evolving, magnetic topologies from 2006 to 2012 and in particular during the intensely observed 3-month period in 2006/7. 

\subsection{Differential Rotation}
\label{SDR}

Surface differential rotation on solar-type stars has been measured using a range of techniques: the Fourier transform of spectral lines \citep[e.g.][]{2006A&A...446..267R}, cross-correlation between images \citep[e.g.][]{2000MNRAS.316..699D}, spot tracking, particularly using $\it{KEPLER}$ data \citep[e.g.][]{2013A&A...557A..11R,2015A&A...583A..65R} and Ca $\textsc{ii}$ H \& K \citep[e.g.][]{1995ApJ...438..269B}. Incorporating a solar-like differential rotation law, as defined in equation~\ref{DR}, into the modelling process has been successfully applied to a number of late F-/early-G stars \citep[e.g.][]{2002MNRAS.334..374P,2004MNRAS.351..826P,2004MNRAS.352..589B,2005MNRAS.357L...1B,2006MNRAS.370..468M}.
\begin{equation}
	\label{DR}	
	\Omega(\theta) = \Omega_{eq} - \Delta\Omega \sin^{2} \theta 
\end{equation}
where $\Omega(\theta)$ is the rotation rate at latitude $\theta$ in \rdd, $\Omega$\subs{eq} is the equatorial rotation rate and $\Delta\Omega$ is the rotational shear between the equator and the pole, both in \rdd. This technique utilizes a grid search for the two differential parameters, by systematically adjusting $\Omega$\subs{eq} and $\Delta\Omega$ and determining $\chi^{2}_r$ for a fixed number of iterations. This grid search produces a non-uniform $\chi^{2}_r$ landscape from which we fit a two-dimensional paraboloid to determine the $\Omega$\subs{eq} - $\Delta\Omega$ combination that best fits the data along with 1-$\sigma$ errors. 

\citet{2008MNRAS.384...77M} demonstrated the possibility of applying differential rotation modelling to relatively long timescales provided that the magnetic topology is relatively stable. Differential rotation measurements were attempted on the complete, 3-month dataset in 2006/7 using the Stokes $\it{I}$ LSD profiles. However, no differential rotation was found. Differential rotation measurements were attempted for the three individual datasets: the CFHT data from November 30 to December 11, 2006, the TBL data from January 25 to February 4, 2007 and the TBL data from February 15 to February 28, 2007. Additionally, differential rotation measurements were attempted on the datasets from 2008 and 2012. Despite these attempts, no measurable result was found due to the inability to find a uniquely located minimum value on the $\chi^{2}_{r}$ landscape. We conclude that, differential rotation is likely (as shown next using Stokes $\it{V}$) although not measurable using Stokes $\it{I}$.

\begin{figure}
\begin{center}
\includegraphics[trim = 5cm 0cm 5cm 0cm, scale=0.80, angle=0]{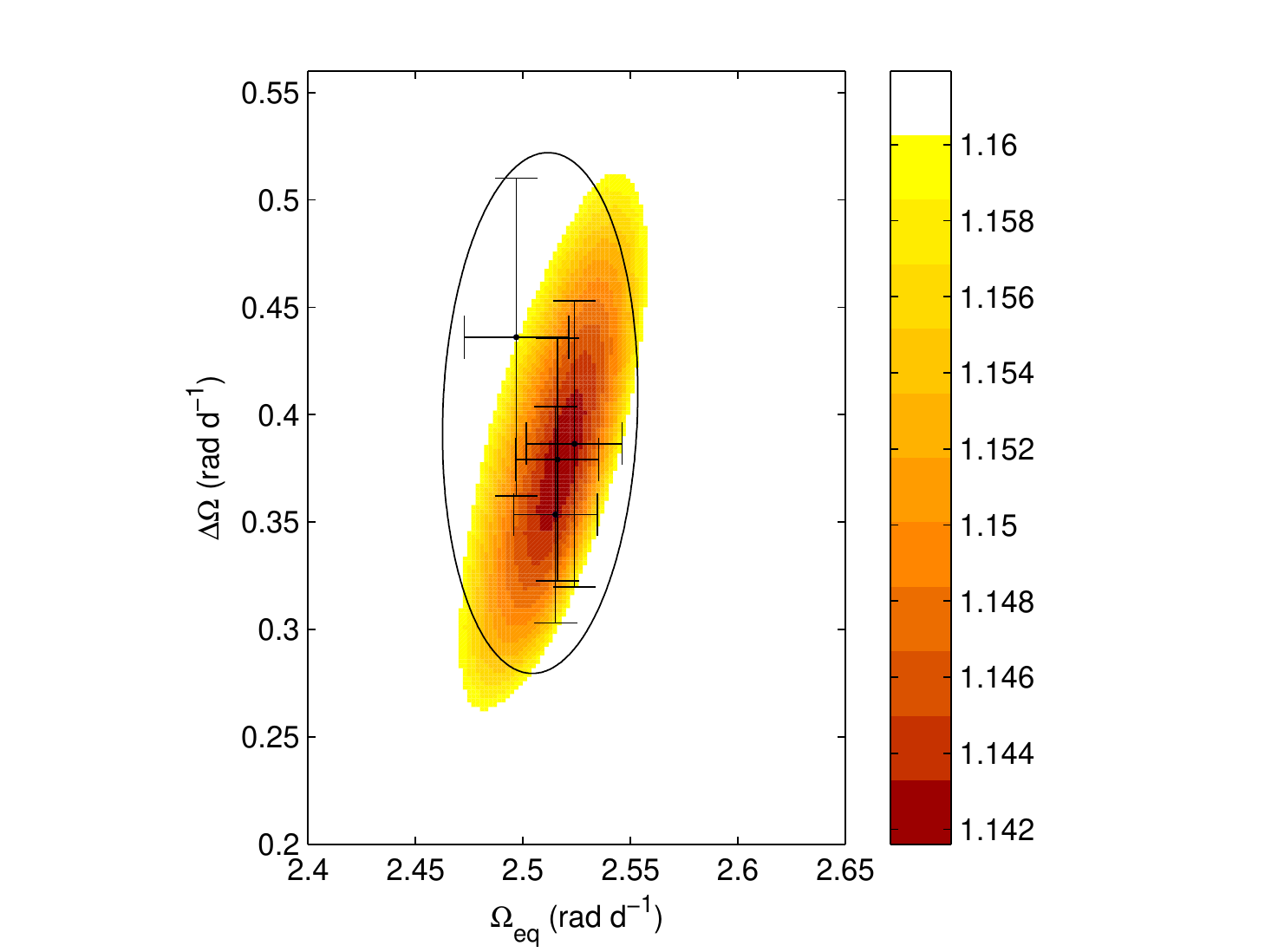}
\caption{Surface differential rotation $\chi^2_r$ minimization landscape for the Stokes $\it{V}$ profiles for EK~Draconis using the dataset from January 2007. The coloured landscape shows the $\chi^2_r$ values obtained systematically changing the $\Omega_{eq}$ and $\Delta\Omega$ pairs for the optimum parameter set. This is a 2-$\sigma$ projection, with the darker colours indicating a lower $\chi^2_r$ value. Superimposed on this landscape is the variation ellipse showing the differential rotation measurements for the Stokes $\it{V}$. This variation ellipse was generated by varying some of the stellar parameters, including the star's inclination ($\pm$~5$\degr$), \vsini\ ($\pm$~0.1~\kms) and the global magnetic field strength ($\pm$~10\%). The error bars on each individual measurement are 1-$\sigma$ errors in the paraboloid fit.}
\label{Showmap_EKDra}
\end{center}
\end{figure}

Like for the brightness maps, $\Omega_{eq}$ and $\Delta\Omega$ pairs were generated using the Stokes $\it{V}$ data and tested using the $\chi^{2}_r$ minimization technique. Whereas the determination of differential rotation was not possible with the Stokes $\it{I}$ data, the Stokes $\it{V}$ data produced differential rotation measurements in all datasets. One hypothesis is that the brightness maps were dominated by one high-latitude feature whereas the magnetic maps have features spread over a range of latitudes. A typical $\chi^{2}_r$ minimization landscape is shown in Fig. \ref{Showmap_EKDra}. This landscape was produced using the optimal parameter set for the January 2007 Stokes $\it{V}$ data. The $\Omega_{eq}-\Delta\Omega$ measurement was found by fitting a two-dimensional paraboloid to the $\chi^{2}_r$ landscape. This determines a 1-$\sigma$ error estimate of that fit. The value of this measurement, coupled with the error estimate, was superimposed on $\chi^{2}_r$ landscape.  

Several more $\chi^{2}_r$ landscapes were produced by systematically varying stellar parameters including the star's inclination angle ($\pm$~5$^{\degr}$), \vsini\ ($\pm$~0.1~\kms) and the global magnetic field strength ($\pm$~10\%) \citep{2002MNRAS.334..374P}. Each minimum $\Omega_{eq}-\Delta\Omega$ pair, with the 1-$\sigma$ error, was calculated and superimposed on this landscape shown in Fig. \ref{Showmap_EKDra}. An ellipse was generated to encompass all of these differential rotation measurements and was used to estimate the overall error in the differential rotation measurement. For this particular (Jan. 2007) dataset, the equatorial rotational velocity, $\Omega_{eq}$ was estimated to be 2.52~$\pm$~0.05 \rdd\ with a rotational shear, $\Delta\Omega$, of 0.38~$\pm$~0.13 \rdd. The error quoted for the differential rotation parameters could be more appropriately called a ``variation'' as it was determined by varying the parameters, as described above, and the associated ellipse known as a ``variation'' ellipse \citep{2011MNRAS.413.1949W}. However, the true error associated with this, and other differential rotation measurements listed in Table \ref{EKDra_Mapping_Table}, may be slightly larger due to intrinsic spot evolution during each data collection time frame \citep[e.g][]{2012A&A...540A.138M}. When considering all available differential rotation measurements across the six years of observations, the average equatorial rotational velocity, $\Omega_{eq}$, was $\sim$2.50~$\pm$~0.08 \rdd\ with an average rotational shear, $\Delta\Omega$, of $\sim$0.27$_{-0.26}^{+0.24}$~\rdd.

\begin{table*}
\begin{center}
\caption{Summary of magnetic topology evolution of EK~Draconis from December, 2006 to January, 2012. The second column shows the global magnetic field strength. The third column is the number of observations ($\phi$). The fourth and fifth columns show the percentage of the total magnetic energy contained in the poloidal and toroidal components. The next three columns show the percentage of the poloidal field that is contained in the dipole ($\ell$=1), quadrupole ($\ell$=2) and octupole ($\ell$=3). The next three columns show the same information of the toroidal field. The final two columns show the percentage of the axisymmetric modes (modes with m = 0) in each of the poloidal and toroidal components. Variations (errors) are based on the systematic recalculation of the models based upon varying stellar parameters: $\Omega_{eq}$, $\Delta\Omega$, \vsini, v$_{rad}$, global magnetic field and inclination.} 
\label{magpot_EKDra}
\begin{tabular}{lcccccccccccc}
\hline
Year & $<|$B$|>$ & no. of & pol. 	  & tor.    & dip.    & quad.   & oct.    & dip     & quad.   & oct.    & axisym.    & axisym. \\
          & G	& of $\phi$	& \% tot. & \% tot. & \% pol. & \% pol. & \% pol. & \% tor. & \% tor. & \% tor. & \% pol.  & \% tor. \\ 
\hline
Dec 2006 & 90$\pm{2}$ & 7 & 17$\pm{1}$  & 83$\pm{1}$ & 41$\pm{3}$  & 16$\pm{2}$  & 14$\pm{1}$ & 90$\pm{2}$ & 7$\pm{2}$  & 1$\pm{1}$  &  35$\pm{3}$ & 99$\pm{0.5}$ \\
Jan 2007 & 81$\pm{2}$ & 7 &  32$\pm{1}$  & 68$\pm{1}$ & 50$\pm{3}$  & 16$\pm{1}$  & 7$\pm{1}$  & 79$\pm{4}$ & 15$\pm{3}$ & 2$\pm{1}$ & 6$\pm{1}$ & 96$\pm{1}$ \\
Feb 2007 & 54$_{-2.5}^{+3.0}$ & 5 & 41$\pm{2}$  & 59$\pm{2}$ & 24$\pm{4}$  & 10$\pm{2}$  & 14$\pm{1}$ & 79$\pm{5}$ & 13$\pm{4}$ & 2$\pm{1}$  & 17$\pm{4}$ & 95$\pm{1}$ \\
Jan 2008 & 59$_{-1.7}^{+2.0}$ & 5 & 37$\pm{1}$ & 63$\pm{1}$ & 33$\pm{3}$	 & 9$\pm{0.5}$   & 16$\pm{2}$ & 82$\pm{4}$ & 14$\pm{4}$ & 2$\pm{1}$  & 8$\pm{1}$ & 97$\pm{1}$\\
Jan 2012 & 92$_{-3.1}^{+3.4}$ & 10 & 43$\pm{2}$  & 57$\pm{2}$ & 7$\pm{0.5}$   & 17$\pm{2}$  & 17$\pm{1}$ & 68$\pm{4}$ & 14$\pm{2}$ & 5$\pm{1}$ & 17$\pm{2}$ & 89$\pm{2}$ \\
\hline
\multicolumn{13}{|l|}{Using fewer profiles of the Jan. 2007 data to match the number of profiles used in Feb. 2007} \\
Jan 2007 & 58$_{-1.7}^{+2.1}$ & 5 & 41$\pm{2}$ & 59$\pm{2}$ & 31$\pm{3}$ & 32$\pm{2}$ & 9$\pm{2}$ & 79$\pm{3}$ & 15$\pm{4}$ & 2.5$\pm{0.3}$ & 13$\pm{4}$ & 96$\pm{0.7}$ \\
\hline
\end{tabular}
\end{center}
\end{table*}

\section{Discussion}

EK~Draconis is a young, very active Sun-like star that underwent significant photospheric and chromospheric changes during our observations. This is particularly evident during the intense time sampling in December 2006, January 2007 and February 2007. 

\subsection{Brightness maps}
\label{DiscussionBrightness}

The brightness maps, in Fig. \ref{StokesI_SB}, were generated from the Stokes $\it{I}$ LSD profiles. EK~Draconis displays spot features at low- to mid-latitudes. This is similar to other young, early G-type stars such as He 699 (\vsinis = 93.5 \kms), in the $\alpha$ Persei cluster \citep[e.g.][]{2001MNRAS.326.1057B,2002MNRAS.331..666J} and HD~141943 (\vsinis  = 35~ \kms) \citep{2011MNRAS.413.1922M}. Additionally,  EK~Draconis displays a relatively large, intermediate-latitude feature at approximately $\sim$40 to 70$\degree$\ latitude in the 2006/7 observing epochs, similar to that observed by \citet{2007A&A...472..887J}. The extensive high-latitude spot feature shown in Fig. \ref{StokesI_SB} appears to coincide with the enhanced chromospheric activity shown in Fig. \ref{fig:Chromospheric_3months}.

This study extends the original 2001 and 2002 observations of \citet{2007A&A...472..887J} until 2012; an interval of approximately ten years. During this ten-year time-frame, the mid- to high-latitude spot features appeared to migrate poleward to form a giant polar spot on EK~Draconis. This is the first time a polar spot has been recorded on this star since the observations made by \citet{1998A&A...330..685S} in 1995, although their analysis at the time was inconclusive regarding the existence of a polar spot. A hypothesis for this poleward migration involves the transport of magnetic flux through the process of meridional circulation \citep[e.g.][]{2001ApJ...551.1099S,1999A&A...344..911K}. It is yet to be determined whether the polar spots on young solar-type stars are formed at high latitudes by the strong Coriolis effect in these rapidly rotating stars, or formed at low latitudes and pushed poleward by subsurface meridional flows. \citet{2005AN....326..287W} have reported tentative evidence for large poleward meridional circulation on giant stars. \citet{2013ApJ...774L..29Z} report poleward meridional flow on the Sun has a speed of 15~\mss. If spot migration is indeed the mechanism, then the drift from 2007 until 2008 was approximately 5$\degree$\ in latitude, indicating that the drift rate would have to be approximately 2.0 \mss. More massive stars such as KIC~8366239 (R = 5.30~$\pm$~0.08~$R_{\sun}$: \citet{2012Natur.481...55B}) have a meridional flow speed of 26 \mss\ or $\sim$13$\degree$\ per year \citep{2012AN....333.1028K} and Arcturus (R = 25.4~$\pm$~0.2 $R_{\sun}$: \citet{2011ApJ...743..135R}) have a meridional flow speed of 170 \mss or $\sim$17$\degree$\ per year \citet{2011AN....332...83K}. Alternatively, if a spot group disappeared and was replaced by another emerging at the surface our maps could signify intrinsic spot evolution on a timescale less than the year between the 2007 and 2008 observations, so the present data cannot be used to decide between spot migration and evolution.

As the Stokes $\it{I}$ data did not enable differential rotation to be measured, a solid-body model was able to fit down to the noise level, with the reduced $\chi^{2}$ values of less than one obtained shown in Table \ref{EKDra_Mapping_Table}. Nevertheless, some systematic residuals remained, and so individual model and observed profiles were normalized by subtracting them from average profiles before being compared in Fig. \ref{EKDra_StokesI}. The advantage of this approach was to accentuate any mis-fitting during the modelling process. This was particularly evident for phases $\phi$~$\sim$~0.26 to 0.3 in the 2012 data. For example, removing the first four profiles (phases $\phi$ = 0.285 to 0.290) had the effect of constraining the significant spot feature at phase $\sim$ 0 to a slightly lower latitude. We attribute the mis-fits between the stellar and modelled profiles as due to differential rotation and spot evolution. Nevertheless, these maps, assuming solid-body rotation, clearly show the location of large intermediate-latitude features during 2006/7 and 2008 with a distinctive polar spot appearing in 2012.

Constraining spot features using small datasets is problematic in Stokes $\it{I}$ mapping. For example, the dataset in late February 2007 and again in January 2008 only had five phases to reconstruct the brightness map. The DI code could not effectively recover both the latitude and phase of these features. This is shown in Fig. \ref{StokesI_SB} with the smearing of the high-latitude feature ($\sim$ 70$\degree$) and the finger-like features crossing several degrees of latitude. For this reason, no DI (or ZDI) was attempted for the January 2009 data as only four phases of observations were recorded.

EK~Draconis' \vsini\ is at the extreme lower limit that permits constructing robust DI brightness maps. For example, \citet{2015MNRAS.449....8W} and \citet{2016MNRAS.459.4325M} unsuccessfully attempted DI on HD~35296 and $\tau$ Bo\"{o}tis respectively, with both stars having a \vsini~$\sim$~15.9 \kms. \citet{2016A&A...593A..35R} produced brightness maps of EK~Draconis using the same data as presented here from 2007 and 2012. Their brightness  maps show different mean spot latitudes when compared to our maps. One reason could be they used the published \vsini~=~16.8 \kms\ \citep{2005ApJS..159..141V} whereas we determined the \vsini~=~16.4 \kms\ using our complete profile fitting approach. The estimation of \vsini\ for slow and moderate rotators is particularly challenging. \citet{1994MNRAS.269..814C} found that over-estimating \vsini\ produces excess equatorial spots while under-estimating \vsini\ produces an excess of high-latitude spot features. This is the reason why we fitted the entire profile instead of merely fitting the wings of the LSD profile. Thus our approach aims to use the observations we have obtained to measure \vsini\ rather than rely on previously published values.  

\subsection{Magnetic maps and configurations}

The magnetic maps in Fig. \ref{StokesV_maps} show a strong, almost unipolar, azimuthal field with complex and varying radial and meridional field structures. This is similar to other BY~Draconis-type stars such as HN~Pegasi \protect\footnote{SpType: G0V, \citet{2006AJ....132..161G}} \citep{2015A&A...573A..17B} and HD~171488 \protect\footnote{SpType: G2V, \citet{1999A&AS..138...87C}} \citep{2006MNRAS.370..468M,2011MNRAS.411.1301J} that show complex and variable magnetic field geometry. During the intensive 3-month investigation, the magnetic field evolved from predominantly toroidal (∼80 per cent) to a more balanced poloidal-toroidal (∼40-60 per cent) field. This reorganization resulted in differing levels of activity observed on EK~Draconis. The Ca $\textsc{ii}$ H \& K and H$\alpha$ spectral lines are formed in the mid-levels of the chromosphere, while the Ca $\textsc{ii}$ IRT lines are formed in the lower levels of the chromosphere. Fig. \ref{fig:Chromospheric_3months} shows the variation in these indices. Variation of all indices due to stellar rotation is evident, as explained in Sect. \ref{chromo}. What is more interesting is that the H$\alpha$ and Ca $\textsc{ii}$ IRT indices hints that EK~Draconis was more active during the January 2007 series of observations when the magnetic field was undergoing a reorganization to a more balanced configuration. This is less obvious when considering the Ca $\textsc{ii}$ H \& K index with the scatter much higher in the data. This could be due to the spectral lines appearing on the edge of the detector where the continuum is not as easily determined. Even removing the overlapping order, as explained in Sect. \ref{chromo} the continuum was not as well-defined as with the other spectral lines. Nevertheless, it may be concluded that the variation in the level of the activity occurring in the lower- to mid-levels of the chromosphere was due to the reorganization of the magnetic field.

\begin{figure}
\begin{center}
\includegraphics[scale=0.57, angle=0]{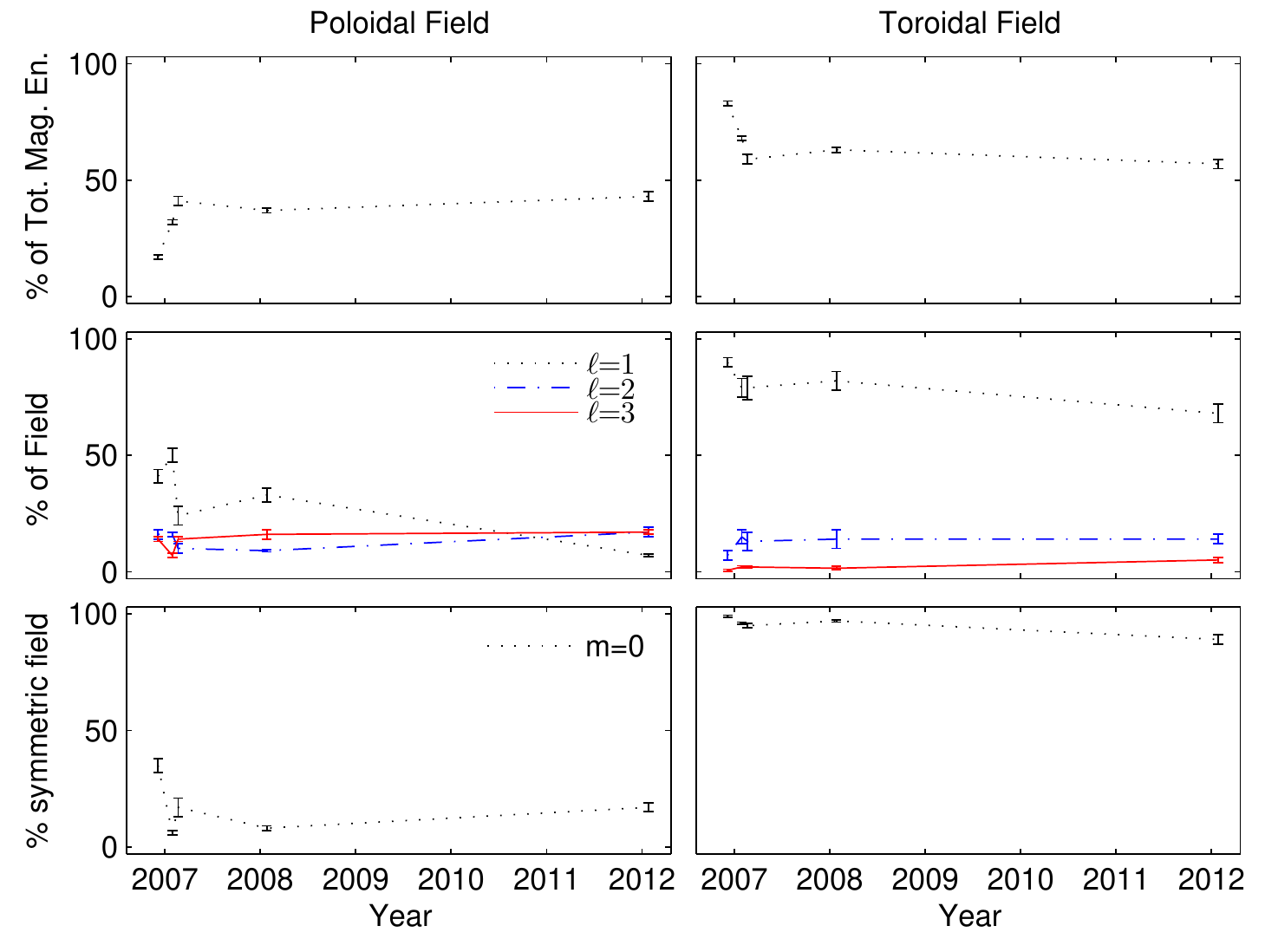}
\caption{The variation in the magnetic field as a function of time. The left series of panels focusses on the poloidal field configuration while the right series of panels focus on the toroidal field configuration. The top two panels shows the variation in the respective field strength as a percentage of the total magnetic field energy. The middle two panels show the various components for dipole ($\ell$ = 1: black $\cdots$), quadrupole ($\ell$ = 2: blue -~$\cdot$~-) and octupole ($\ell$ = 3: red ---). The bottom two panels show the percentage of the poloidal (left) and toroidal (right) field that are axisymmetric (m = 0). The error bar on each datapoint was generated by varying some of the stellar parameters, including the star's $\Omega_{eq}$, $\Delta\Omega$, inclination ($\pm$~5$\degr$), \vsini\ ($\pm$~0.1~\kms) and the global magnetic field strength ($\pm$~10\%). }
\label{EKDra_poloidal}
\end{center}
\end{figure}

The dynamo operating in the Sun is known as a $\alpha-\Omega$ dynamo \citep[e.g.][]{1955ApJ...122..293P,2010LRSP....7....3C}. The $\alpha$-effect is associated with the twisting of the magnetic fields by helical convection while the $\Omega$-effect produces the toroidal field by shearing the poloidal field due to differential rotation. \citet{2011A&A...535A..48H} argue that the $\alpha$-effect also has a role to play in the production of the toroidal field as opposed to the toroidal field being produced entirely by the $\Omega$-effect. In the Sun, the tachocline, or shear, layer is at the base of the convection zone \citep{2003ARA&A..41..599T}; hence the toroidal component is in the sub-surface layers of the Sun. However, this might not be the case on young Sun-like stars. The toroidal component of the large-scale dynamo field may manifest itself in the form of strong azimuthal magnetic fields on, or near, the surface of rapidly-rotating solar-type stars \citep[e.g.][]{2003MNRAS.345.1145D,2004MNRAS.351..826P}. \citet{2010ApJ...711..424B} used 3-dimensional magnetohydrodynamic modelling in their anelastic spherical harmonic code to show that persistent wreathes of azimuthal magnetic fields can be produced. The azimuthal magnetic field observed on EK~Draconis is practically unipolar and strongly negative in all epochs although the January 2007 map does show some small and very weak positive regions in this configuration as shown in Fig. \ref{StokesV_maps}. This may be due to the reorganization from a strongly toroidal field ($\sim$83$\pm$1 per cent) to a weaker toroidal configuration ($\sim$59$\pm$2 per cent) as shown in Fig. \ref{EKDra_poloidal}. The error listed in these values and in Fig. \ref{EKDra_poloidal} and Table \ref{magpot_EKDra} are based on the systematic recalculation of the models based upon varying stellar parameters: $\Omega_{eq}$, $\Delta\Omega$, \vsini, v$_{rad}$, global magnetic field and inclination. However, the true error bars could be larger due to the intrinsic evolution of the magnetic field \citep[e.g.][]{2012A&A...540A.138M}. On the longer time scale of five years, there appears to be no polarity reversals as the strongly persistent azimuthal field remains strongly negative, with some changes in the radial and meridional field observed, but again, no evidence for a polarity reversal (see Fig. \ref{StokesV_maps}). It could be that the magnetic cycle of EK~Draconis is longer than five years or that EK~Draconis has yet to establish significant polarity reversals, unlike the much older $\tau$ Bo\"{o}tis \citep{2008MNRAS.385.1179D,2009MNRAS.398.1383F,2016MNRAS.459.4325M}. Further observations are required to determine which is true.

\citet{2009ARA&A..47..333D} suggest that stars with a Rossby number of $\le$ 1, but more massive than 0.5~$M_{\sun}$, has a substantial toroidal component with a mostly non-axisymmetric poloidal component. The Rossby number \citep{1984ApJ...279..763N} for EK~Draconis was determined using:

\begin{equation}
\label{Rossby}
	\ R_{o} = \frac{P_{obs}}{\tau_{c}}
\end{equation}
where $\tau_{c}$ is the convective turnover time. $\tau_{c}$ was determined using the empirical formula of \citet{2011ApJ...743...48W} and is shown in equation \ref{T_C}:
 
\begin{equation}
\label{T_C}
	\ \log \tau_{c} = 1.16 - 1.49 \log\frac{M}{M_\odot} - 0.54 \log^{2}\frac{M}{M_\odot}
\end{equation}

equation \ref{T_C} holds for stellar masses in the range of $0.09~\le~M/M_{\sun}~\le~1.36$.  EK~Draconis has a convective turnover time of 12.6~d; hence a Rossby number of $\sim$ 0.2. EK~Draconis conforms to this pattern as it has a substantial toroidal component, even though it was reorganizing itself from the strongly toroidal field of 83~$\pm$~1 per cent in December, 2006, to 59~$\pm$~2 per cent by the end of February, 2007. This toroidal field remained firmly axisymmetric during the five years of observations, with over 90 per cent of the toroidal field being axisymmetric, as shown in the bottom panels of Fig. \ref{EKDra_poloidal}. This is consistent with the findings by \citet{2015MNRAS.453.4301S} that strong toroidal fields are predominantly axisymmetric. The poloidal field, as shown in Fig.~\ref{EKDra_poloidal}, has increased in dominance from $\sim$17 per cent in December, 2006, to $\sim$41 per cent by the end of February, 2007. The poloidal field was predominantly non-axisymmetric reaching a maximum of 82~$\pm$~3 per cent during the dataset from January 2007. These observations of significant toroidal fields  with non-axisymmetric poloidal fields during the five years of observations support the conclusions of \citet{2009ARA&A..47..333D} for stars that have Rossby numbers less than one but are more massive than 0.5 $M_{\sun}$. 

Our magnetic maps are broadly consistent with the magnetic maps produced by \citet{2016A&A...593A..35R} of EK~Draconis using the data taken in 2007 and again in 2012, although their phase is aproximately 0.5 different from ours. Whereas the brightness maps are sensitive to the \vsini\ measurement, the magnetic maps are not as sensitive hence our maps appear very similar to those of \citet{2016A&A...593A..35R}. Their maps show a strong, unipolar azimuthal magnetic field with complex and varying radial and meridional field structures. Additionally, we have incorporated differential rotation into the imaging process. Differential rotation will be explored in more detail in Sect. \ref{DiscussionDR}.

The February 2007 and January 2008 magnetic maps were reconstructed using only five observations.  \citet{2011MNRAS.413.1939M} found that small datasets provided difficulties in determining the magnitude and makeup of the global mean magnetic field. To examine the effect of using fewer profiles in the imaging process, we reduced the number of profiles used in the reconstruction of the dataset from January 2007 to the number in the dataset from late-February. Additionally, we attempted to approximate the appropriate phase coverage of the dataset. This resulted in a reduction of the mean field strength from 81~G to 58~G. The conclusion is that the observed variation in the mean field strength from the January to February 2007 is most likely a result in the reduced number of profiles, as opposed to any actual changes in the mean magnetic field strength of the star itself. When considering the configuration of the field, the poloidal component was slightly enhanced when using fewer profiles (32~$\pm$~1 to 41~$\pm$~2 per cent), as shown in Fig. \ref{EKDra_poloidal} and listed in Table \ref{magpot_EKDra}. Additionally, the toroidal field remained predominantly axisymmetric and the poloidal field remained predominately a complex, non-axisymmetric field. Thus we can conclude that the global magnetic field topology has evolved with a weakening of the toroidal component during the three month time-frame.

\subsection{Latitude dependence of the magnetic field}

\begin{figure}
\begin{center}
\includegraphics[trim=3.5cm  0cm 5cm 0cm, scale=0.68, angle=0]{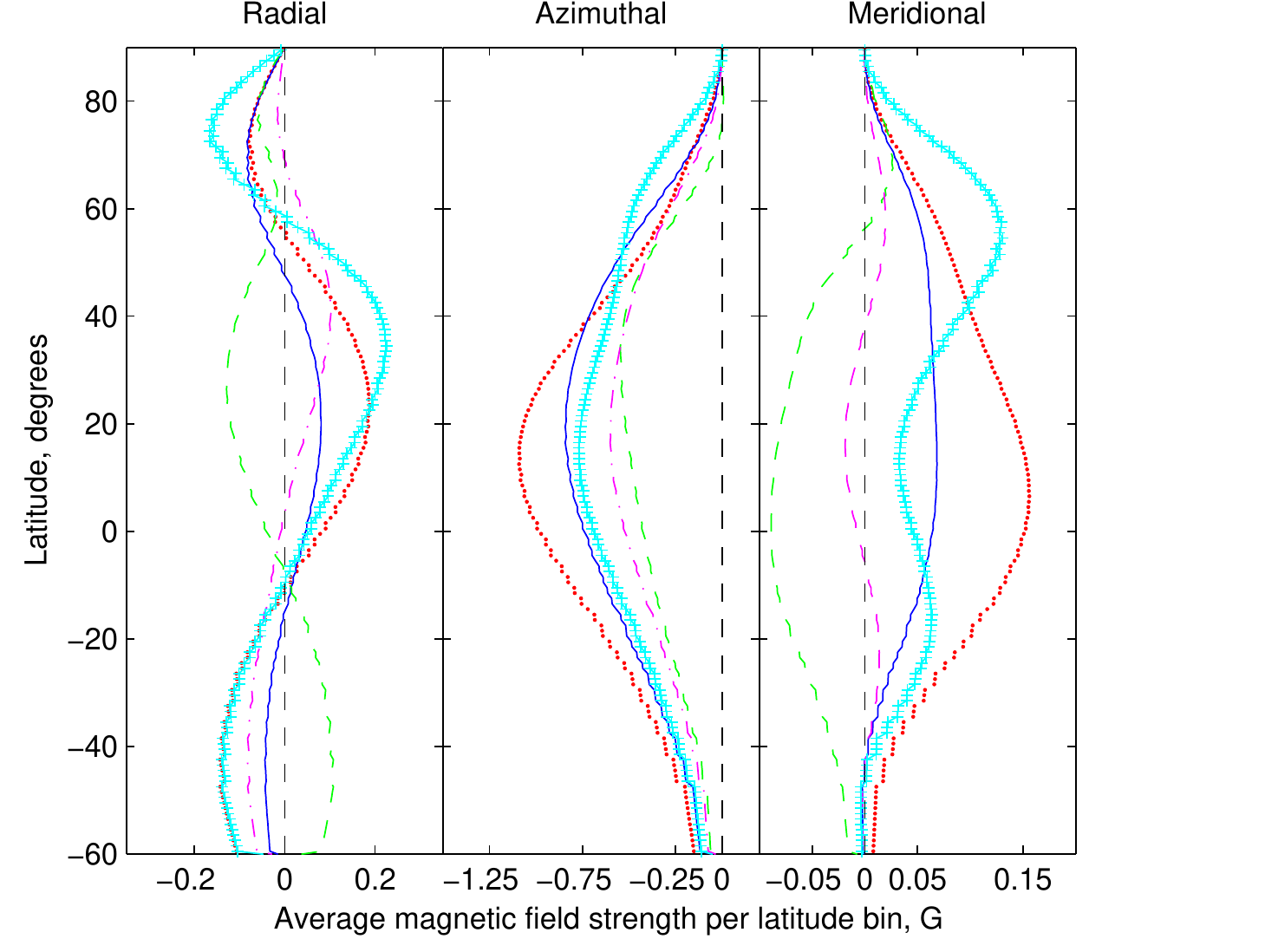}
\caption{This shows a comparison of the average magnetic field strength (G) as a function of latitude for the radial field (left), azimuthal field (middle) and meridional field (right) for the five epochs observed. The dots (red) represent the fractional magnetic fields for the CFHT run, the solid (blue) line represent the data from the TBL in January while the dashed (green) line represents the field from late February, 2007. The dot-dash (magenta) line is for 2008 while the $\circ$ (cyan) line is for the 2012 data.}
\label{EKDra_B_Vs_Latitude_all}
\end{center}
\end{figure}

Determining the fractional magnetic field strength as a function of latitude enables further quantitative analysis of the distribution of the magnetic fields on the surface of the star. We calculate the average magnetic field per latitude bin, $\it{B_f}(\theta$), as:

\begin{equation}
	\label{V_spottedness}	
	B_{f}(\theta) = \sum_{i=1}^{n} B_{i}(\theta) \frac{\cos(\theta)d\theta}{2}
\end{equation}
where $\theta$ is the latitude and $\it{d}\theta$ the width of the bin. Fig.~\ref{EKDra_B_Vs_Latitude_all} shows the average magnetic field as a function of latitude for each field component. 

The magnetic maps shown in Fig. \ref{StokesV_maps} clearly show a strong, almost unipolar azimuthal magnetic field. This is reflected in Fig. \ref{EKDra_B_Vs_Latitude_all} with a stable, strongly negative field at all latitudes, with this field dominating the equatorial regions. The radial and meridional fields appears to have minor changes in latitude distribution during the five-year time-frame. The 2012 meridional field appears to dominate at higher latitudes; perhaps reflecting the poleward migration of the large spot group observed in the brightness maps. There can be some cross-talk between the radial and meridional components, particularly at low latitudes \citep{1997A&A...326.1135D, 2012A&A...548A...8R, 2013A&A...550A..84K}, although with an inclination angle of 60$\degree$\ this is less likely for EK~Draconis.  

\subsection{Differential rotation}
\label{DiscussionDR}

A key driver of the dynamo operating in young Sun-like stars is differential rotation. Applying a solar-like differential rotation to the Stokes $\it{V}$ information, the equatorial region of EK~Draconis was found to rotate at $\Omega_{eq}$ $\sim$2.50~$\pm$~0.08 \rdd, equating to a period of $\sim$ 2.51~$\pm$~0.08~d. Considering the size of the respective error ellipse, as shown in Fig.~\ref{Showmap_EKDra}, the results are consistent with the observations of \citet{2003A&A...409.1017M} who showed that EK~Draconis exhibited solar-like surface differential rotation with the equator rotating faster than the poles. Table \ref{EKDra_Mapping_Table} shows the variation in the differential rotation measurement. Temporal variation in differential rotation measurements have been noted in the literature, including the photometric studies listed already. Additionally, longitudinal DI studies of the K-dwarf star AB~Doradus \protect\footnote{SpType: K0V, \citet{2006A&A...460..695T}} has observed variations in the laptimes (the time it takes for the equator to lap the polar regions) from $\sim$70 to $\sim$140~d \citep{2002MNRAS.329L..23C,2003MNRAS.345.1187D}. The variations in the rotational shear observed on EK~Draconis are potentially real; however, these variations could be due to the paucity of observations, particularly in the late February, 2007 data. The level of differential rotation is much higher for the dataset from January 2007 and again the dataset from January 2008 when compared with the other epochs. Numerical simulations have been performed by \citet{2002MNRAS.334..374P} that show phase coverage and/or observational cadence can effect the magnitude of the $\Omega_{eq}$ - $\Delta\Omega$ recovered. For EK~Draconis, observational cadence was certainly a factor with the lower rotational shear during the dataset from February 2007. However, the CFHT dataset has similar cadence to January 2007 and again in 2008 but showed significantly lower $\Omega_{eq}$ (2.42~$\pm$~0.1 \rdd\ compared with 2.52~$\pm$~0.05 \rdd\ using January 2007 data) and $\Delta\Omega$ (0.19~$\pm$~0.18 \rdd\ compared with 0.38~$\pm$~0.13 \rdd). One can speculate that the strong toroidal field of the CFHT dataset might have masked the differential rotation signature. The dataset from January 2007 and again from 2008 had a more balanced field configuration with both datasets showing a poloidal field of $\sim$37 and 38 percent respectively. The rotational shear in both 2007 and 2008 epochs was $\sim$0.39~\rdd. Further observations would be required to determine the exact nature of the relationship between the magnetic configuration and the recovered differential rotation. Nevertheless, the conclusion is that EK~Draconis has a significant rotational shear that could be as high as 0.39 \rdd, equating to a laptime of $\sim$16 days.

The error in the $\Delta\Omega$ values impact on the laptime calculated and listed in Table \ref{EKDra_Mapping_Table}. As explained in Sect. \ref{SDR}, the error in $\Delta\Omega$ was found by varying certain parameters (\vsini, inclination angle, $<$B$_{mod}>$) and determining the minimum $\chi_{r}^{2}$ in each of the respective landscapes. The individual error bar on each measurement, as shown in Fig. \ref{Showmap_EKDra}, was a 1-$\sigma$ error in the fit to the 2-dimensional paraboloid. On some occasions, such as the differential rotation measurement with the CFHT data, the 1-$\sigma$ error bar is relatively large thereby increasing the size of the error ellipse.

\subsection{Moderate rotators}

EK~Draconis is very similar to HD~35296 and HD~29615 that were studied by \citet{2015MNRAS.449....8W} (see Paper I). All three of these Sun-like stars have similar age, mass, radius and rotational velocity. Table \ref{HD35296_HIP21632_EKDra} highlights these similarities. All exhibit reasonably high levels of differential rotation, with laptimes (where the equator laps the polar regions) ranging from $\sim$13~d for HD~29615 to $\sim$29~d for HD~35296. 

\begin{table*}
\begin{center}
\caption{A comparison between the fundamental parameters of HD~35296, HD~29615 and EK~Draconis:} 
\label{HD35296_HIP21632_EKDra}
\begin{tabular}{llll}
\hline
Parameter 				& HD~35296$^{a}$ & HD~29615$^{a}$	  & EK~Draconis 	\\
\hline
Equatorial period, d 		   	& 3.5~$\pm$~0.2  & 2.34~$\pm$~0.2 	  &  2.51~$\pm$~0.07 	\\
Inclination angle, $$\degr$$	   	& 65~$\pm$~5 	 & 70~$\pm$~5 		  &  60~$\pm$~5   	\\
Photospheric Temperature, T$_{phot}$, K 	& 6080	 & 5820~$\pm$~50	  &  5561 		\\
$\Delta$Temp = T$_{phot}$ - T$_{spot}$, K 	& --	& 1900 		  	  &  1700  		\\
Stellar Radius: $R_{\sun}$		& $\sim$1.21	&  $\sim$1.0		  &  $\sim$0.94 	\\ 
Stellar Mass:   $M_{\sun}$		& $\sim$1.10 	&  $\sim$0.97  		  &  $\sim$0.95		\\
Projected Rotational Velocity, \vsini, \kms  & 15.9~$\pm$~0.1 & 19.5~$\pm$~0.1    & 16.4~$\pm$~0.1 	\\
$\log g$					& 4.31~$\pm$~0.03 & 4.43~$\pm$~0.04 	  & 4.47~$\pm$~0.08     \\
Convective turnover time, d		& 10.3		& 16.0			  & 17.2		\\
Rossby Number				& 0.34		& 0.15			  & 0.15		\\
Stokes $\it{V}$: $\Omega_{eq}$, in \rdd & 1.804~$\pm$~0.005   & 2.74$_{-0.04}^{+0.02}$  & $\sim$2.50~$\pm$~0.08 \\
Stokes $\it{V}$: $\Delta\Omega$, in \rdd & 0.22$^{+0.04}_{-0.02}$   & 0.48$_{-0.12}^{+0.11}$ & 0.27$^{+0.24}_{-0.26}$ 	\\ 
laptime$^{b}$, d		&  $\sim$29   &  $\sim$13	  & $\sim$23 		\\
\hline
\end{tabular}
\end{center}
$^{a}$ data taken from paper I, \citet{2015MNRAS.449....8W}, and the references therein. \\
$^{b}$ The laptime is the time it takes for the equator to lap the polar regions. \\
\end{table*}

All three stars are considered as moderately rotating (MR) solar-type stars \citep[as defined by][]{2011PASA...28..323W}, with all these stars having projected rotational velocities (\vsini) ranging from 5 to 20 \kms. It is within this range where the strength of the magnetic dynamo is believed to be related to the star's rotation rate \citep{2014MNRAS.441.2361V}. Above the rotational velocity of $\sim$20 \kms\ the strength of the magnetic dynamo is believed to be no longer dependent on stellar rotation. All three of these Sun-like stars, with rotation rates below 20 \kms, exhibit near-surface azimuthal fields indicating that the dynamo could be fundamentally different from that operating in the Sun today. The toroidal fields of all three Sun-like stars are strongly axisymmetric while the poloidal fields tend to be more non-axisymmetric; however, HD~29615 does not follow the other two stars where the poloidal field is predominately axisymmetric as only $\sim$33 per cent of the poloidal energy held in the non-axisymmetric configuration. The Rossby number, a measure of how strongly the Coriolis force is capable of affecting the convective eddies, is lowest in HD~29615 amongst the three stars ($\it{Ro}$~=~0.14). \citet{2009ARA&A..47..333D} suggests that small $\it{Ro}$ values indicate very active stars rotating fast enough to ensure that the Coriolis force strongly impacts convection. One can speculate that the star with the lowest Rossby number also has the highest differential rotation shear, as is the case when comparing these three Sun-like stars. However, for very low Rossby numbers, (R$_{o}~\le$~0.1), in the case of M-dwarf stars, the differential rotation is very small \citep{2005MNRAS.357L...1B,2008MNRAS.390..545D}. \citet{2014MNRAS.438L..76G}, using 3-D simulations, concluded that maximum $\Delta\Omega/\Omega$ is achieved at moderate R$_{o}$ (R$_{o}$~\textless~1.0), and then converges towards zero at the fastest rotation rates (see Fig. 4 of their work). 

Surface differential rotation has a key role in the generation of the stellar magnetic field. Surface differential rotation measurements have shown variations when using brightness features compared with the value found using magnetic features \cite[e.g.][]{2003MNRAS.345.1187D,2015MNRAS.449....8W}. One hypothesis is that these features are anchored at different depths within the convective zone, meaning that the radial differential rotation structure must be very different from that observed on the Sun. HD~29615 is one star that shows this variation, albeit an extreme example with $\Delta\Omega$ = 0.48$_{-0.12}^{+0.11}$ using Stokes $\it{V}$ and 0.07$_{-0.03}^{+0.10}$ using Stokes $\it{I}$. In the absence of a definitive measurement using DI, we cannot confirm this hypothesis with EK~Draconis.

\section{Conclusions} 

Our investigations have observed strong differential rotation using the magnetic features on EK~Draconis. Additionally our observations show significant evolution occurring on EK~Draconis in a short three-month period. These changes are possibly the result of the magnetic field reorganizing itself from a strongly toroidal field ($\sim$80 per cent) to a more balanced poloidal-toroidal field ($\sim$40-60~per cent) in only three months, as shown in Fig.~\ref{EKDra_poloidal}. The poloidal field on EK~Draconis was predominately non-axisymmetric while the toroidal field as almost entirely axisymmetric during all epochs. This is consistent with other Sun-like stars. EK~Draconis appeared to show intermediate-latitude features during earlier epochs while a distinctive polar spot was observed during 2012. Additionally, a persistent, almost unipolar azimuthal field was observed at all epochs indicating that no polarity reversals were found in our data.

\section*{Acknowledgements}

Thanks must go to the staff of the CFHT and TBL in their assistance in taking these data. The authors thank the anonymous referee for their insightful comments that has significantly enhanced this paper. The authors thank Professor Jardine, Dr See, Ms Boro Saikia and Mr Mengel on their valuable contributions to this work. This research has made use of NASA's Astrophysics Data System. This research has made use of the VizieR catalogue access tool, CDS, Strasbourg, France. The original description of the VizieR service was published in A\&AS 143, 23. This research used the High Performance Computing Facility at the University of Southern Queensland.

\bibliography{bibliography}


\label{lastpage}

\end{document}